\documentstyle[12pt,aaspp4]{article}
\begin{document}
\parindent=1.0cm

\font\twelverm=cmr12
\font\ninerm=cmr9
\def\HII{\hbox{\twelverm H\thinspace\ninerm II}}

\title{Near-Infrared Adaptive Optics Imaging of the Central Regions of Nearby 
Sc Galaxies. II. NGC 247 and NGC 2403}

\author{T. J. Davidge \altaffilmark{1} $^{,}$ \altaffilmark{2}}

\affil{Canadian Gemini Office, Herzberg Institute of Astrophysics,
\\National Research Council of Canada, 5071 W. Saanich Road,\\Victoria,
B.C. Canada V9E 2E7\\ {\it email:tim.davidge@nrc.ca}}

\author{S. Courteau \altaffilmark{1}}

\affil{Herzberg Institute of Astrophysics,
\\National Research Council of Canada, 5071 W. Saanich Road,\\Victoria,
B.C. Canada V9E 2E7}

\begin{center}
and
\end{center}

\affil{Department of Physics \& Astronomy, University of British 
Columbia, \\Vancouver, B. C. Canada V6T 1Z1\\ {\it email:courteau@astro.ubc.ca}}

\altaffiltext{1}{Visiting Astronomer, Canada-France-Hawaii Telescope 
Corporation, which is run by the National Research Council of Canada, the 
Centre National de la Recherche Scientifique, and the University of Hawaii.}

\altaffiltext{2}{Guest User, Canadian Astronomy Data Centre, which is operated 
by the Herzberg Institute of Astrophysics, National Research Council of Canada.}

\begin{abstract}

	$J, H,$ and $K'$ images obtained with the Canada-France-Hawaii 
Telescope adaptive optics system are used to investigate the star-forming 
histories of the central regions of the Sc galaxies NGC 247 and NGC 2403. 
The brightest resolved red stars within 15 arcsec of the nucleus of each galaxy 
are red supergiants (RSGs), indicating that the central few hundred parsecs of 
these galaxies experienced star formation within the last $\sim 0.1$ Gyr. 
While there are indications of galaxy-to-galaxy 
differences in the recent ($t \leq 0.1$ Gyr) star-forming histories, a 
comparison of the $K$ luminosity functions of stars near the centers of NGC 
2403 and M33 indicate that, when averaged over Gyr time scales, the 
star-forming histories of the inner disks of these galaxies have been 
remarkably similar. This is consistent with suggestions that the long-term 
evolution of disks is defined by local characteristics such as mass density. 
It is demonstrated that NGC 247 and NGC 2403, like M33, harbour nuclear star 
clusters with stellar contents that differ from the surrounding central light 
concentrations, which in turn have near-infrared spectral energy 
distributions that are similar to old stellar systems. The nucleus 
of NGC 2403 is significantly bluer than that of the other two galaxies. If the 
nuclei of NGC 247, NGC 2403, and M33 are subject to similar amounts of 
extinction then this indicates that NGC 2403 harbours the youngest nuclear star 
cluster in the sample. The $K-$band surface brightnesses near the centers of 
NGC 247 and NGC 2403 are 1 -- 2 mag arcsec$^{-2}$ lower than in M33. Finally, 
it is noted that young or intermediate-age nuclear star clusters are a common 
occurence in nearby spirals, indicating that nuclear star formation in these 
objects is either continuous or episodic on time scales of $0.1 - 1$ Gyr. This 
is consistent with models that have been proposed to explain the properties of 
the Galactic Center. 

\end{abstract}

\keywords{galaxies: individual (NGC 247, NGC 2403) - galaxies: evolution - galaxies: stellar content - galaxies: bulges - galaxies: nuclei}

\section{INTRODUCTION}

	The bulges of spiral galaxies are not monolithic entities that 
evolve in isolation; rather, the evolution of these systems is influenced by 
their environment. As a reservoir containing significant quantities of gas, the 
disk might be expected to play a significant role in bulge evolution, and the 
large-scale structural characteristics of spiral galaxies provide evidence 
supporting a bulge-disk connection, in the sense that the impact of the disk 
on bulge properties becomes more important towards progressively later 
morphological types (Andredakis, Peletier, \& Balcells 1995; Courteau, de Jong, 
\& Broeils 1996). Interactions with the disk could explain the differences 
in central structural properties noted between the bulges of spirals spanning a 
range of morphological types (e.g. Phillips et al. 1996; Courteau 1997), as 
well as the presence of sub-structures (Carollo, Stiavelli, \& Mack 1998) and 
nearly exponential light profiles (Courteau 1997; Carollo 1999; 
MacArthur et al. 2002) in many bulges. Physical processes 
that can trigger the motion of disk material into the central regions of 
galaxies include bar instabilities (Friedli \& Benz 1993, 1995; Sellwood \& 
Moore 1999), galaxy-galaxy interactions (Barnes \& Hernquist 1992, Mihos \& 
Hernquist 1996), and dynamical friction (Noguchi 1999, 2000).

	The bulges of spiral galaxies earlier than Sc appear to be dominated by 
old stars, suggesting that any recent interactions with the disk (1) have been 
modest, and (2) have not affected the basic properties of these systems. The 
colors of bulges in the Hubble Deep Field indicate that these objects
formed at high redshift (Abraham et al. 1999), while the 
tight relation in the $(B-I, I-H)$ diagram defined by the bulges of nearby 
early-type spirals is consistent with both a modest age spread and an old age 
(Peletier et al. 1999). The spectral energy distributions (SEDs) of the bulges 
of M31 and M81, which are the two closest intermediate-type spirals, 
indicate that these systems are dominated by stars that formed during early 
epochs (Davidge 1997, 2001b; Bica, Alloin, \& Schmidt 1990; Kong et al. 
2000). These observations are consistent with 
galaxy formation simulations, which predict that the main bodies of 
bulges form at $z > 2$ (e.g. Katz 1992; Steinmetz \& Mueller 1994; 
Moll\'{a}, Ferrini, \& Gozzi 2000; Gnedin, Norman, \& Ostriker 2000), although 
star formation may continue into more recent epochs (Moll\'{a} et al. 2000).
The collisionless accretion of satellites can 
also reproduce the observed structural characteristics of bulges, and 
these models also predict relatively old bulges, as the accretion must 
predate the formation of the thin disk (Aguerri, Balcells, \& Peletier 2001). 

	The Milky-Way is a barred (e.g. Blitz \& Spergel 1991; Dwek et al. 
1995) intermediate-type (Blanco \& Terndrup 1989) spiral galaxy. 
The Galactic bulge is of fundamental importance as an age calibrator 
since individual main sequence stars can be resolved. The Galactic bulge has a 
global mass-to-light ratio that is similar to the M31 bulge (Kent 1992), and so 
the bulges of the Galaxy and M31 ought to have similar stellar contents. After 
accounting for contamination from stars in the intervening disk, the main 
sequence turn-off of the Galactic bulge is indicative of an old age (Feltzing 
\& Gilmore 2000). This is consistent with studies of 
bulge globular clusters (Minniti 1995), which have 
ages of at least $\sim 10$ Gyr (Ortolani et al. 1995; Chaboyer et al. 2000), 
although metallicity-dependent calibration issues frustrate efforts to 
determine when these clusters formed with respect to clusters in the 
outer halo (Rosenberg et al. 1999). Finally, further constraints on the age 
of the Galactic bulge can be obtained from the age of the Galactic disk, as the 
disk is expected to be one of the last components of the Galaxy to form. 
Using the kinematic properties of main sequence and sub-giant branch stars 
selected using Hipparcos information, Binney, Dehnen, \& Bertelli (2000) find 
that the disk has an age of $11.2 \pm 0.75$ Gyr, and is only $\sim 1$ Gyr 
younger than the bulk of globular clusters.

	Although the main bodies of bulges in early and intermediate-type 
spiral galaxies are dominated by old stars, the innermost regions of these 
objects contain young or intermediate-age stars, indicating that star formation 
continued to recent epochs. There are young star clusters near the Galactic 
Center (GC; e.g. Cotera et al. 1996), and these are in 
the process of being tidally disrupted (Figer et al. 1999). 
The youngest stars near the GC have solar-like Fe abundances, and a 
metallicity distribution function that differs from that of older stars in 
the bulge, but is similar to that of solar-neighborhood giants 
(Ramirez et al. 2000), ostensibly suggesting -- but not conclusively proving 
-- that they formed from gas that originated outside the bulge. 
While it is problematic whether or not the young clusters found near 
the GC formally belong to the bulge, they nevertheless contribute young 
stars to the inner bulge as they dissolve 
(Figer et al. 1999). In the case of M31, the 
absence of bright resolved sources at ultraviolet wavelengths 
(Bohlin et al. 1985, Brown et al. 1998) indicates that the central regions of 
this galaxy do not contain the very young stars that are seen near the GC, 
although the integrated spectrum of the central regions of the M31 bulge 
shows signatures of an intermediate-age component (Bica et al. 1990; Davidge 
1997; Sil'Chenko, Burenkov, \& Vlasyuk 1998), which may be the relic of nuclear 
star forming activity that terminated a few Gyr in the past (Davidge 2001b).

	The relative contribution from any young or intermediate-age component 
to the total light from a bulge might be expected to increase towards later 
morphological types if, as suggested by Andredakis et al. (1995) and 
Courteau et al. (1996), interactions with the disk become more significant for 
these systems. Late-type spiral galaxies are thus important laboratories for 
assessing interactions between the inner disk and bulge. In the first paper of 
this series (Davidge 2000; hereafter Paper 1), deep $J, H$, and $Ks$ 
images obtained with the Canada-France-Hawaii Telescope (CFHT) Adaptive 
Optics Bonnette (AOB) were used to investigate the stellar content of the 
central regions of M33, which is the closest Sc galaxy. These data indicate 
that the star forming history in the inner disk of M33 during recent epochs has 
not been continuous. Moreover, the depth of $2.3\mu$m CO absorption in the 
integrated light of the central light concentration of M33 is consistent 
with a moderately low metallicity, indicating that it could not have formed 
from material in the present-day disk. While the low metallicity suggests 
that the main body of the central light concentration is likely 
not young, and may even be an inward extension of the halo (Bothun 
1992), the nucleus has a near-infrared SED 
reminiscent of intermediate age Magellanic Cloud clusters and Sgr A (Davidge 
2000). Gordon et al. (1999) concluded that the nucleus of M33 contains a 
significant amount of dust, with a major burst of star formation occuring some 
70 Myr in the past.

	It remains to be determined if the stellar content in the central 
regions of M33 is representative of other Sc galaxies. In the present work, the 
stellar contents in the central regions of the Sc galaxies NGC 247 and NGC 
2403 are investigated and compared with M33. These galaxies are close enough 
that the brightest inner disk stars can be resolved. Moreover, they 
are viewed at intermediate inclination angles, so that their central 
light concentrations are likely not heavily obscured by dust in the disk. 
Finally, gas in the disk of NGC 2403 has a chemical 
composition that is similar to M33 (Garnett et al. 1997), and so 
young and intermediate age stars in these galaxies should have comparable 
metallicities.

	The morphological type, integrated apparent $B$ brightness, total 
extinction, true distance modulus, and integrated absolute $B$ brightness of 
M33, NGC 247, and NGC 2403 are listed in Table 1. The extinction estimates are 
based on the model described by Tully \& Fouqu\'{e} (1985), 
as applied by Pierce \& Tully (1992). The extinction in 
the disks of spiral galaxies can be patchy and, in the case of NGC 2403, A$_V$ 
towards \HII\ regions varies between 0.2 -- 0.9 mag (McCall, Rybski, \& 
Shields 1985; Petersen \& Gammelgaard 1996), which corresponds to E(J--K) 
ranging from 0.03 mag to 0.15 mag using the Rieke \& Lebofsky (1985) 
extinction curve. Consequently, while the extinction entries in Table 1 
likely hold for each system when averaged over large angular scales, 
they may not reflect the actual values in the inner disk and nucleus, 
and we estimate the uncertainty in E(J--K) to be $\pm 0.06$ mag based on 
the extinction measurements towards \HII\ regions in NGC 2403.

	The distance moduli of M33 and NGC 2403 in Table 1 are 
from van den Bergh (1992) and Freedman \& Madore (1988), respectively. 
NGC 247 has by far the most uncertain distance modulus, in large part because 
there are no published Cepheid observations. The distance modulus listed in 
Table 1 for this galaxy was derived from the Tully-Fisher relation, using line 
widths and integrated $B$, $R$, and $I$ brightnesses 
from Pierce \& Tully (1992), and the calibrating relations 
from Sakai et al. (2000). The result agrees within a few 
tenths of a dex with the distance derived for NGC 253, which is located 
close to NGC 247 on the sky, using the brightest halo (Davidge \& Pritchet 
1990) and disk (Davidge, Le F\`{e}vre, \& Clark 1991) stars.

	The paper is structured as follows. The observations and data 
reduction procedures are described in \S 2. The photometric properties of 
resolved stars in the inner disks of NGC 247 and NGC 2403 are discussed in \S 
3, while the integrated photometric properties of the central light 
concentrations of these galaxies are compared in \S 4. Comparisons are made 
with M33 throughout the paper using the data discussed by Davidge (2000), which 
have been processed to simulate the appearance of this galaxy if observed at 
the same distance and with the same angular resolution as NGC 2403. 
A summary and discussion of the results follows in \S 5.

\section{OBSERVATIONS AND REDUCTIONS}

	The data were recorded with the CFHT AOB (Rigaut et al. 1998) and KIR 
imager during observing runs in 1998, 1999, and 2000. The detector in KIR is a 
$1024 \times 1024$ Hg:Cd:Te array, with each pixel subtending 0$\farcs$034 
$\times$ 0$\farcs$034 on the sky; hence, a $35 \times 35$ arcsec region is 
imaged.

	$J, H,$ and $Ks$ images centered on an uncatalogued $K = 10$ 
star, located 23 arcsec south west of the nucleus of NGC 2403 
with $\alpha$ = 07:36:48 and $\delta$ = $+$65:35:41 (E2000), 
were recorded on the night of UT March 8, 1998. This star was the reference 
beacon for AO compensation, and is not listed in 
either the HST or USNO Guide Star catalogues because 
the central regions of NGC 2403 are overexposed on the POSS. Ten 30 
second integrations per filter were recorded at each corner of a 0$\farcs$5 
$\times 0\farcs$5 square dither pattern, so that the total exposure time was 
$10 \times 30 \times 4 = 1200$ sec per filter. This same dither pattern was 
adopted for the other fields discussed in this paper. The FWHMs of stars in 
the final images of this field are 0$\farcs$20 ($J$), 0$\farcs$15 ($H$), and 
0$\farcs$17 ($Ks$). This field will be refered to as the `deep' NGC 2403 field 
for the remainder of the paper. 

	$J, H,$ and $Ks$ images of a second NGC 2403 field, which includes 
the galaxy nucleus and will be refered to as the `central' NGC 2403 field, were 
recorded during the night of UT September 10, 2000. The field center was 
chosen so that there would be significant overlap with the deep field, and the 
uncatalogued $K = 10$ star described in the previous paragraph was used 
as the reference source for AO guiding. Two 60 second exposures per filter 
were recorded at each dither position, and the total exposure time is thus $2 
\times 60 \times 4 = 480$ seconds per filter. The FWHMs of stars in the final 
images of this field are 0$\farcs$17 ($J$), 0$\farcs$15 ($H$), and 
0$\farcs$19 ($Ks$). 

	$J, H,$ and $Ks$ images of the center of NGC 247 were recorded during 
the nights of UT June 20 and 21 1999. The NGC 247 line of sight 
passes near the South Galactic Pole, and there are no stars near the center 
of this galaxy that are bright enough ($R \leq 15$) to serve as AO guide 
sources; consequently, the central light concentration of NGC 247 was used as 
the AO reference source. The NGC 247 data were recorded with 
60 second integration times, and the total exposure time was 8 
minutes in $J$, 17 minutes in $H$, and 20 minutes in $Ks$. The image quality of 
these data is poor by AOB standards, as the central light concentration is 
extended and near the faint limit of the AOB wavefront sensor. The 
FWHMs of stars in this field are 0$\farcs$68 ($J$), 
0$\farcs$54 ($H$), and 0$\farcs$48 ($Ks$).

	The data were reduced using the procedures described in Paper 1, 
which correct for the dark current, pixel-to-pixel sensitivity variations, 
vignetting along the optical path, the DC sky level, interference fringes, and 
thermal emission signatures. The processed images were then aligned, 
median-combined, and trimmed to the area of common overlap for each field. The 
final $Ks$ images are shown in Figures 1, 2, and 3; the central 
light concentrations of NGC 247 and NGC 2403 are marked in Figures 2 and 3.

\section{PHOTOMETRIC PROPERTIES OF RESOLVED STARS}

\subsection{Photometric Measurements}

	A single PSF was constructed for each field using tasks in DAOPHOT 
(Stetson 1987), and the brightnesses of individual stars were then measured 
with the PSF-fitting routine ALLSTAR (Stetson \& Harris 1988). While 
anisoplanicity causes the PSF to change with distance from the 
AO reference star, photometric studies of globular clusters with the AOB 
indicate that these variations typically introduce scatter of only a few 
hundredths of a magnitude over the KIR field (e.g. Davidge \& Courteau 1999; 
Davidge 2001a). The unresolved background in each galaxy, which can cause 
scatter in the photometry if it is non-uniform over angular scales 
comparable to, or smaller than, that of the sky annulus used by ALLSTAR, 
was subtracted from each field using the iterative technique described by 
Davidge et al. (1991).

	Artificial star experiments, in which scaled PSFs were added with noise 
to the images with the DAOPHOT ADDSTAR routine, were used (1) to estimate 
completeness, (2) to assess the uncertainties introduced by random noise and 
crowding, and (3) to measure systematic effects in the photometry that cause 
stars near the faint limit to appear brighter than they actually are. The 
artificial stars were assigned colors representative of real objects in each 
field, and were constructed using the PSFs discussed in the preceding 
paragraph. The artificial star experiments thus assume that anisoplanicity does 
not significantly affect the photometry, and evidence to support this 
assumption is given in \S 3.2.

\subsection{Color-Magnitude and Two-Color Diagrams}

	The $(H, J-H)$, $(K, H-K)$ and $(K, J-K)$ CMDs of stars detected in the 
NGC 247 and NGC 2403 fields in all three filters are shown in Figures 4, 5, 
and 6. The PSF wings of the bright AO guide star elevate the sky level in the 
region immediately surrounding this object; not only does this 
affect the faint limit, but it also adds noise, which may be mis-identified 
as stars. To avoid these problems, sources within 4$\farcs$25 of this 
star were not considered in this study.

	The CMDs contain a red plume, the width of which is 
consistent with the scatter predicted by the artificial star 
experiments. This indicates that observational errors, rather than 
intrinsic star-to-star differences in photometric properties, are the 
dominant source of scatter in the data. The good agreement 
between the predicted and observed widths of the red plume also indicates 
that anisoplanicity is not a major contributor to the photometric errors.

	The photometric calibrations of the two NGC 2403 datasets are 
remarkably consistent, as the mean difference in the $K$ brightness of sources 
common to both fields is $\Delta K = -0.01 \pm 0.02$, where the difference is 
in the sense center field minus deep field. The quoted 
uncertainty is the standard error of the mean. Some of 
these stars might be variable; however, the modest dispersion 
in $\Delta K$ suggests that the brightest sources common to both fields 
have typically not varied in $K$ by more than a few hundredths of a 
magnitude in the 1.5 years separating the observations.

 	The CMDs in Figures 4 - 6 are dominated by 
a mixture of red supergiants (RSGs) and stars evolving on the AGB. RSGs are 
massive stars that burn material in a non-degenerate 
core, while AGB stars are burning material in a shell around an electron 
degenerate core (e.g. Iben 1974; Wood, Bessell, \& Fox 1983). At solar 
metallicities the onset of the AGB in simple stellar systems occurs 
at an age log(t$_{Gyr}$) = 8.1, which 
corresponds to stars near 5 M$_{\odot}$ (Girardi et al. 2000). 

	RSGs and AGB stars have very different near-infrared 
photometric properties. RSGs in the LMC and SMC define a roughly vertical 
sequence on the (M$_K, J-K)$ CMD, which extends from M$_K = -9$ to M$_K = -12$, 
whereas long period variables (LPVs), the majority 
of which are evolving on the AGB, are redder, fainter, 
and distributed over a broader range of colors (e.g. Figure 8 of Davidge 1998, 
which was constructed using data from Elias, Frogel, \& Humphreys 1985, Wood 
et al. 1983, and Wood, Bessell, \& Paltoglou 1985). RSGs dominate star 
counts in the LMC and SMC when M$_K < -10$, and so the red stars with $K < 
17.5$ in NGC 2403 and $K < 17$ in NGC 247 are likely RSGs. AGB stars dominate 
in the Magellanic Clouds when M$_K > -9$, which corresponds 
to $K > 18.5$ in NGC 2403 and $K > 18$ in NGC 247. The 
tendency for the NGC 2403 CMDs to become redder when $K > 18$ 
is consistent with the onset of a bright AGB component.

	The CMDs of the two NGC 2403 fields also show that a modest blue 
population, with $K \sim 18$, is present. These stars have M$_K \sim -9.5$, and 
are too bright to be main sequence OB stars, which have M$_K > -5$ (e.g. Cotera 
et al. 2000); rather, these stars are likely early-type supergiants. In 
addition, two of the brightest sources near the center of NGC 2403 are 
non-stellar, and these objects also appear to be extended in the HST F160W 
and F187N images of NGC 2403 used by B\"{o}ker et al. (1999) in their study of 
the central regions of nearby spiral galaxies. These objects have unremarkable 
brightnesses in the F187N images, indicating that they have weak or 
non-existent Pa$\alpha$ emission; hence, they 
are likely not compact \HII\ regions, but could be 
blends of two or more stars, or possibly even compact star clusters. 
These objects are marked in Figure 2.

	The presence of red and early-type supergiants indicates that star 
formation occured within the past $\sim 0.1$ Gyr in the inner disk of NGC 2403.
Drissen et al. (1999) conducted an H$\alpha$ survey 
of NGC 2403, and an inspection of their Figure 1 indicates that while there is 
diffuse H$\alpha$ emission near the center of the galaxy, the 
surface brightness is markedly lower than in the surrounding areas. 
Massive stars can deposit significant amounts of energy into the 
interstellar medium (e.g. Leitherer, Robert, \& Drissen 1992), and the 
resulting winds can produce \HII\ bubbles (e.g. Tarrab 1983). 
The distribution of H$\alpha$ emission near the center of NGC 2403 suggests 
that the recent star-forming event near the center of NGC 2403 may have been 
sufficiently intense to form such a bubble.

	The (M$_K, J-K$) CMDs of NGC 247 and NGC 2403, corrected for 
interstellar extinction using the entries listed in Table 1 and the Rieke \& 
Lebofsky (1985) reddening law, are shown in Figure 7. Also plotted is a CMD of 
the central regions of M33. NGC 247 and NGC 2403 are more distant than M33, 
and this could affect efforts to make comparisons with M33, as 
blending, which occurs when stars fall 
within the same resolution element, will be less of a concern for nearby 
systems. To remove this potential source of systematic error, the M33 data from 
Paper 1 were processed to simulate the appearance of this galaxy if (1) 
shifted to the same distance as NGC 2403, and (2) observed with the same 
angular resolution as the NGC 2403 central field. This was done by 
block-averaging each M33 image in $4 \times 4$ pixel groups, and then smoothing 
the result with a gaussian to match the FWHM of the NGC 2403 data. The 
M33 measurements shown in Figure 7 were obtained from the resulting images. 

	In Paper 1 it was found that the brightest stars near the center of M33 
have M$_K \sim -9$, and this is roughly consistent with the CMD in Figure 7. 
Nevertheless, there is a spray of objects with M$_K \leq -9$ that are likely 
blends introduced by the simulated increase in distance. The relative number of 
blended sources in the NGC 2403 data will be smaller than predicted 
from the M33 data, as only a region that is very close to the central 
light concentration is sampled in M33, whereas the NGC 2403 
data covers a much larger range of distances from the central light 
concentration, with a corresponding lower mean stellar density.

	The red plume of NGC 2403 in Figure 7 peaks near M$_K 
\sim -11.5$, which is comparable to the peak RSG brightness for galaxies with 
integrated brightnesses comparable to NGC 2403 (Rozanski \& Rowan-Robinson 
1994). The brightest source in the two NGC 2403 fields is stellar, although 
the second and third brightest sources are non-stellar, and hence are not 
individual stars. The spatial distribution of stars within two magnitudes of 
the peak $K$ brightness near the center of NGC 2403 (i.e. with M$_K \leq 
-9.5$) is shown in Figure 8, and it is evident that recent star-forming 
activity, as traced by these RSGs, has occured mainly to the South of the 
nucleus.

	The peak brightnesses in NGC 247 
and M33 are roughly one magnitude fainter in $K$ than in NGC 2403. Although 
having similar peak brightnesses, the stellar contents of the central regions 
of NGC 247 and M33 differ, in that there is a more-or-less continuous 
population of stars in the interval between M$_K = -9.5$ and $-10.5$ in the 
CMD of the former, whereas this brightness interval is poorly populated in the 
CMD of the latter.

	When compared with NGC 2403, the angular resolution 
of the NGC 247 data are relatively poor, and blending of unresolved stars 
is a potential concern. The effects of blending were 
investigated using the F160W NIC3 image (frame N4K416C9Q) of
NGC 247 that was obtained as part of the survey of the central 
regions of nearby late-type galaxies described by B\"{o}ker et al. (1999). 
The image was retrieved from the CADC HST archive and 
pipeline processed; cosmic rays and bad pixels were identified and 
suppressed by interpolating between surrounding pixels. The processed 
image was gaussian smoothed to an angular resolution 
of $0\farcs6$ to match the CFHT NGC 247 observations. 
DAOPHOT and ALLSTAR were then used to measure the brightnesses of sources 
from the smoothed and unsmoothed images, and the difference between these, 
$\Delta$, was computed. For sources between $H = 17$ and $H = 20$ 
$\Delta \sim 0$, with a standard deviation $\pm 0.05$ mag. $\Delta$ does not 
depart significantly from 0 until $H \geq 20$, which is well 
below the faint limit of the CFHT data. Hence, the 
photometric measurements of bright sources in NGC 247 
shown in Figures 4, 5, and 6 are not affected by blending.

	M33 is much closer than the other galaxies, and so the KIR field 
samples an intrinsically smaller portion of this galaxy. This difference in 
spatial coverage could affect the comparisons in 
Figure 7 if there is a radial gradient in stellar content, 
or simply because of small number statistics. Indeed, McLean \& Liu (1996) 
found stars as bright as M$_K = -10.5$ within a few arcmin of the 
center of M33. If only the portion of NGC 2403 that corresponds to the region 
sampled in M33 is considered, then the brightest star has M$_K = -10.2$, 
while the brightest star in the corresponding region of NGC 247 has 
M$_K = -9.5$. Thus, significant differences in bright stellar content remain 
even when similar regions of the galaxies are compared.

	The $(J-H, H-K)$ two-color diagrams (TCDs) of the stars 
detected in the central fields of NGC 247 and NGC 2403 
are shown in Figure 9, along with fiducial sequences defined by Milky-Way 
RSGs and LPVs in the LMC and SMC. Aside from a modest clump of early-type 
supergiants with $H-K = J-H = 0$ in the NGC 2403 TCD, 
the data for both galaxies match the RSG and LPV fiducial sequences. 
There is an absence of extremely red LPVs in the upper right hand corner of 
the TCD. These stars tend to have M$_K \geq -8.5$, and hence are too faint to 
be detected with the current data. 

\subsection{Luminosity Functions}

	The completeness-corrected M$_K$ luminosity functions (LFs) of stars 
detected in both $H$ and $K$ in NGC 2403 and M33 are compared in Figure 10. The 
LF for NGC 2403 was constructed by combining the LFs of the central and deep 
fields, such that the entries for M$_K > -10$ are from the deep field after 
scaling to match the number of stars in the central field with M$_K < -10$, 
while those for brighter stars are from the central field. The resulting 
composite LF was then scaled to match the mean surface brightness in the M33 
field using the $r-$band surface brightness profiles for M33 and NGC 2403 
measured by Kent (1987a,b). Despite the differences in stellar content evident 
when M$_K \leq -9.5$ in Figure 7, the LFs of M33 and NGC 2403 agree within 
their estimated uncertainties between M$_K = -9.5$ and M$_K = -8$, which is a 
brightness interval dominated by luminous AGB stars. Thus, the star-forming 
histories of the innermost disks of NGC 2403 and M33 have been similar on 
$\sim 1$ Gyr time scales. 

\section{NUCLEAR PROPERTIES}

	In \S 3 it was demonstrated that the recent ($t \leq 1$ Gyr) star 
forming histories of the inner disks of NGC 247, NGC 2403, and M33 have been 
different. If star forming episodes in the nuclei and inner disks of late-type 
spiral galaxies are somehow related (e.g. they are 
triggered by the same mechanism), then the nuclear 
photometric properties of NGC 247, NGC 2403, and M33 ought to show significant 
differences as well. The nuclear photometric properties of these galaxies 
are investigated in this section. 

\subsection{Color Profiles}

	The colors of the central light concentrations of NGC 247 and NGC 2403 
were measured in a series of concentric circular apertures, and the results are 
plotted in Figures 11 (integrated measurements) and 12 (differential 
measurements). The minimum aperture diameter for NGC 2403 was set at 
$0\farcs$25, and this was increased in $0\farcs$25 increments. A coarser 
sampling was used for NGC 247 because of the poorer angular resolution of these 
data: the minimum aperture diameter was $0\farcs$5, and this was increased in 
$0\farcs$5 increments. The measurements, which were restricted to a maximum 
radius of 1$\farcs$25, were made after the $H$ and $K$ data for each galaxy had 
been smoothed to match the angular resolution of the $J$ data, which had 
the poorest angular resolution in each dataset. The local 
background was measured in an annulus immediately surrounding each central 
light concentration to minimize contamination from the inner disk. 
A corresponding set of aperture measurements were made from the 
M33 data from Paper 1 after these were block-averaged and 
gaussian-smoothed to match the distance and angular resolutions of the NGC 247 
and NGC 2403 data.

	While the means exist to perform a more complicated 
photometric analysis of the central regions of these galaxies with 
finer angular sampling, it was decided not to pursue these 
because anisoplanicity introduces subtle non-uniformities 
in the PSF, which could have a major impact on 
photometric measurements made on angular scales less than the FWHM. 
Rather, it was felt that aperture measurements with angular scales comparable 
to the FWHM of the images would provide a more robust means of 
comparing the central photometric properties of these galaxies. 

	The colors measured for NGC 247 are broadly consistent with those 
published by Frogel (1985), who found that $J-K = 0.75$ and $H-K = 0.14$ 
through a 6$\farcs$3 aperture; for comparison, the current data give $J-K = 
0.68$ and $H-K = 0.13$ through a 1$\farcs$0 radius aperture. Given that the 
uncertainty in the current colors due to the photometric calibration is $\pm 
0.04$ mag, then the agreement with the Frogel (1985) measurements is good.

	Pierini et al. (1997) measured $H-K = 0.17$ 
within 1$\farcs$6 of the center of NGC 2403, and $H-K = 0.14$ within 
4$\farcs$8, while the present data give $H-K = 0.04$ mag within a 1$\farcs$25
radius. We have confidence in the photometric calibration of our NGC 2403 data 
because (1) the locus of resolved stars on the CMDs of M33 and NGC 2403 in 
Figure 7 have similar colors, even though the integrated colors of the central 
light concentrations differ at roughly the $5\sigma$ level, and (2) when 
placed on the near-infrared TCD, stars in NGC 2403 
fall along fiducial sequences (Figure 9), which would not be the case if 
the photometric calibration were greatly in error.

	 The colors in Figures 11 and 12 were also checked using 2MASS images 
of NGC 247 and M33; unfortunately, the NGC 2403 data could not be checked in 
this way since this portion of the 2MASS survey 
has not yet been released. The 2MASS images of NGC 247 give 
a color of $J-K = 0.66$ within 1 arcsec of the galaxy center, which is 
in excellent agreement with what is measured here. The 2MASS data also 
indicate that $J-K$ becomes bluer with increasing radius in NGC 247, 
in qualitative agreement with the color profiles in Figures 11 and 12. 
As for M33, the 2MASS images confirm the relatively red central color of 
this galaxy, although they give $J-K = 0.90$ within the central 
$1\farcs$0 of this galaxy, whereas the current data give $J-K = 0.81$. 
The 2MASS data also indicate that $J-K$ increases towards larger radius in M33, 
in qualitative agreement with the trends in Figures 11 and 12.

	It is evident from Figures 11 and 12 that (1) the nuclear $J-K$ color 
of NGC 2403 is bluer than M33, and (2) the central light concentration of NGC 
2403 contains a marked color gradient, similar in amplitude to that in M33. 
Given the qualitatively similar radial color 
gradients in the central light concentrations of M33 and 
NGC 2403, and that the former contains
a young nuclear star cluster (Gordon et al. 1999; Paper 1), then 
a young nuclear component is likely also present in 
the latter. NGC 2403 contains the bluest nuclear population of the three 
galaxies considered here, while the resolved stellar content within 20 arcsec 
of the nucleus indicates that the inner disk of this galaxy has experienced 
the most recent star-forming activity. This consistency in the relative ages 
of the nucleus and inner disk hints that the star-forming histories of these 
components may have been coupled. 

	The $J-K$ colors of NGC 247 and M33 do not differ significantly when 
$r \leq 0\farcs$5, suggesting that these galaxies have similar nuclear 
stellar contents. It is interesting that the $J-K$ gradient near the center of 
NGC 247 is in the opposite sense to the gradients in M33 and NGC 2403. 
Nevertheless, the near-infrared SED changes 
with radius in a way that is still consistent with an increase in the number of 
younger stars towards smaller radii (see below).

	The extinction in the nucleus is a significant source of uncertainty 
when interpreting Figures 11 and 12. Gordon et al. (1999) find that A$_V 
\sim 1.4$ mag ($\tau_{V} = 1.3$) in the nucleus of M33, which corresponds to 
E(J--K) = 0.23 using the Rieke \& Lebofsky (1985) reddening law. For 
comparison, the extinction listed for M33 in Table 1, which was adopted for 
Figures 11 and 12, gives E(J--K) = 0.04 mag. The effect of adopting the higher 
extinction for the nucleus of M33 is shown in Figures 11 and 12, and it is 
evident that the nuclei of M33 and NGC 2403 may have similar intrinsic colors 
if they are subject to different amounts of extinction, in the sense that M33 
is very dusty and NGC 2403 is relatively dust-free.

	The $(J-H, H-K)$ TCD provides a quantitative means of comparing the 
near-infrared SEDs of the centers of these galaxies. 
Luminosity-weighted $J-H$ and $H-K$ colors were calculated for each galaxy 
in the radial intervals $r \leq 0\farcs$5 and $0\farcs$5 $\leq r \leq 
1\farcs$25, and the near-infrared TCDs of the results
are shown in Figure 13. The NGC 247 and NGC 2403 data were de-reddened using 
the A$_B$ entries in Table 1; however, to demonstrate the effects of 
extinction on the interpretation of the TCD, the M33 data were de-reddened 
using both the A$_B$ entry for this galaxy in Table 1, and the extinction 
estimated by Gordon et al. (1999), and so the TCDs in Figure 13 show 
two points for M33. Also plotted in Figure 13 are (1) the 
colors of Galactic globular clusters from Table 5A of Brodie \& Huchra (1990), 
(2) the colors of confirmed SWB type 1 and 2 Magellanic Cloud clusters, 
which are among the youngest clusters in the Magellanic Clouds (e.g. Hodge 
1983), from Tables 4 and 5 of Persson et al. (1983), (3) the $6\farcs$6 
aperture measurements of the nuclei of nearby Sc galaxies from Table 4 of 
Frogel (1985), and (4) the effects of varying age from 1.5 to 3.0 Gyr for a 
solar metallicity population (top panel), and of varying the metallicity of 
a 1.5 Gyr population from solar to [Fe/H] = $-0.22$ and $+0.25$ (lower panel), 
as predicted from the models computed by Worthey (1994). Frogel (1985) did not 
correct for extinction internal to the galaxies, and his data in 
Figure 13 have been corrected for extinction within each disk by assuming that 
$E(B-V) = 0.09$, as predicted by Tully \& Fouqu\'{e} (1985) for disk galaxies 
with an inclination angle of 45$^\circ$.

	The luminosity-weighted colors in the two radial intervals of each 
galaxy define significantly different locations on the TCD, as expected if 
there are population gradients near the centers of these galaxies. Moreover, 
it is evident from the M33 datapoints in Figure 13 that the 
effects of extinction can be significant. Nevertheless, broad conclusions can 
still be drawn from the TCD.

	NGC 247 and M33 have comparable locations on the TCD when $r \leq 
0\farcs$5, and this is consistent with these systems having similar $J-K$ 
colors in Figures 11 and 12. The innermost regions of NGC 247 and M33 
occupy the same area of the near-infrared TCD as SWB Type 1 and 
2 clusters in the Magellanic Clouds, and this holds for the 
two A$_V$ values considered for M33. The data thus indicate that 
the nuclei of these galaxies are dominated 
by intermediate age populations. It is also evident 
from the top panel of Figure 13 that the nucleus of 
NGC 2403 has a near-infrared SED that is very different from the nuclei of 
the other galaxies. The Worthey (1994) models suggest that the 
location of the nucleus of NGC 2403 on the near-infrared TCD can not 
easily be explained with respect to NGC 247 and M33 simply by involking 
differences in age and/or metallicity. However, 
Worthey (1994) did not model systems with $t \leq 1.5$ Gyr, and the trends 
defined by the models shown in Figure 13 may not hold for significantly 
younger populations.

	The circumnuclear regions of the three galaxies also show a broad range 
of near-infrared photometric properties. The NGC 247 and 
NGC 2403 measurements when $r \geq 0\farcs$5 are located in the same 
region of the near-infrared TCD as Galactic globular clusters; hence, the 
circumnuclear regions of these galaxies have near-infrared photometric 
properties that are consistent with them being old stellar 
systems. In addition, even though NGC 247 has a color gradient that is in the 
opposite sense to that in the other 2 systems, the near-infrared SED is 
still consistent with a radial gradient in mean age, in the sense of 
younger stars occuring at smaller radii.

\subsection{$K$--Band Surface Brightness Profiles}

	The $K-$band surface brightness profiles of the central regions of NGC 
247, NGC 2403, and M33, obtained from the 
aperture measurements discussed in \S 4.2, are compared in 
Figure 14. After adjusting for differences in seeing and distance, the mean 
$K-$band surface brightnesses near the centers of NGC 247 and NGC 2403 are 
$1 - 2$ mag arcsec$^{-2}$ fainter than in M33. This difference in surface 
brightness is not due to uncertainties in the image quality, as experiments 
indicate that even if the image quality of the M33 data were increased by 
0$\farcs$1 with respect to the other systems then the mean surface 
brightness of M33 within the central aperture is lowered by only 
$\sim 0.1$ mag arcsec$^{-2}$. The surface brightnesses are also 
not sensitive to extinction, as adopting the higher extinction for 
the M33 nucleus proposed by Gordon et al. (1999) changes the $K-$band 
surface brightness near the center of this galaxy by only 0.1 mag, and this is 
in the sense of making the nucleus even brighter. Hence, when comparing mean 
surface brightnesses within a given aperture, M33 has an 
intrinsically higher $K-$band surface brightness than within $0\farcs$12 of 
the center of NGC 2403, and within $0\farcs25$ of the center of NGC 247. 

	The extreme compact nature of the M33 nucleus has long been recognized 
(e.g. Kormendy \& McLure 1993). We suggest that the high central surface 
brightness may not be due exclusively to a high mass concentration, but could 
be, at least in part, a result of stellar content effects. The nucleus of M33 
has a relatively strong central CO index (Paper 1), suggesting that there 
is a centrally concentrated population of red stars, which would elevate the 
$K-$band surface brightness. High angular-resolution 
narrow-band CO images do not yet exist for NGC 247 and NGC 2403, but if future 
observations reveal that the nuclei of NGC 247 and 2403 have weaker CO indices 
than M33 then stellar content will have to be considered when comparing the 
central near-infrared surface brightnesses of these galaxies. 
Based on the $J-K$ colors in Figures 11 and 12 it might be anticipated 
that the central CO indices of NGC 247 and M33 are similar. 

\section{DISCUSSION \& SUMMARY}

	$JHK$ images obtained with the CFHT AOB have been used to investigate 
the stellar contents in the central 30 arcsec of the nearby Sc galaxies NGC 247 
and NGC 2403. Stars as faint as M$_K = -8.5$ are detected in the 
inner disks of both systems. The majority of the resolved stars are RSGs 
or are evolving on the AGB, although early-type supergiants 
have also been detected near the center of NGC 2403.

	M33, NGC 247, and NGC 2403 were included in the broad and narrow-band 
NICMOS imaging survey of late-type systems discussed by B\"{o}ker et al. 
(1999), and their data identified isolated (`I') or diffuse nebulosity (`DN') 
in the $51 \times 51$ arcsec NICMOS3 field of view of each system, indicating 
that recent star formation has occured in the inner disks of these galaxies. 
Our data, which probe a broader range of ages than is possible with emission 
line information, reveal that there have been significant galaxy-to-galaxy 
variations in the recent star forming histories near the centers of these 
galaxies.

	Stochastic variations in the recent star-forming histories of 
fields within a galaxy are to be expected, and Davidge (1998) found 
significant galaxy-to-galaxy variations in the infrared-bright stellar content 
of the late-type Sculptor spiral galaxies NGC 55, NGC 300, and 
NGC 7793. It might be anticipated that such variations in local star-forming 
history will diminish as stars sampling progressively longer time intervals 
are included in the analysis, and the comparison of the $K$ LFs of upper AGB 
stars in the inner disks of these galaxies in Figure 10 indicates that 
the star-forming histories of the inner disks of M33 and NGC 2403 have been 
similar when averaged over Gyr timescales.

	Bell \& de Jong (2000) used optical and near-infrared photometry to 
investigate the stellar contents of nearby galaxies, and concluded that disk 
evolution is driven largely by local mass density. This can be explained if the 
star-forming history depends on the local gas surface density, as 
originally proposed by Schmidt (1959). Given that (1) 
M33 and NGC 2403 are located in relatively uncrowded environments, and hence 
have likely evolved in relative isolation, and (2) the inner disks of these 
galaxies have very similar surface brightnesses (Kent 1987a, b), 
then the Bell \& de Jong (2000) results predict that they would 
have had similar star forming histories, which is 
consistent with the comparison in Figure 10.

	We also compared the integrated near-infrared photometric 
properties of the central regions of NGC 247 and NGC 2403 with those of M33. 
B\"{o}ker et al. (1999) found that NGC 247, NGC 2403, and M33 lack measureable 
nuclear Pa$\alpha$ emission, suggesting an absence of significant nuclear 
star-forming activity during the last $10 - 100$ Myr. An absence of 
line emission could occur in the presence of a young population if the ISM 
has been lost due to winds. However, this can not be the case 
in M33, as the central regions of this galaxy contain dust 
(Gordon et al. 1999), indicating that an ISM is present. 

	There is a $J-K$ color gradient 
within 1 arcsec of the nucleus of NGC 2403, which is in the same sense as 
that in the corresponding region of M33. Given that M33 
contains a young or intermediate-age nuclear population (Gordon et al. 1999; 
Paper 1), which causes a color gradient near the center of that galaxy, 
it is reasonable to conclude that the nucleus of NGC 2403 
also contains a young population. The nuclei of 
NGC 2403 and M33 may have different ages, as the nucleus of 
NGC 2403 has a bluer $J-K$ color than M33, although this conclusion 
assumes that the nuclear dust contents of these systems are similar. The 
near-infrared SED of the central light concentration surrounding the nucleus 
of NGC 2403 is consistent with it being an old simple stellar system, 
at least as defined by Galactic globular clusters. 
Spectroscopic observations will be necessary to determine if the central 
light concentrations of M33 and NGC 2403 have similar metallicities and ages.

	Drissen et al. (2000) used WFPC2 images to investigate the central 
regions of NGC 2403, and measured a half light radius of 3.4 parsecs, or 
0$\farcs$22, for the nucleus. Based on the measured color ($B-V = 
0.78$), and the absence of radio (Turner \& Ho 1994) and 
H$\alpha$ emission, Drissen et al. suggest that the central light 
concentration of NGC 2403 is old. With an integrated color of $J-K = 0.6$ at 
0$\farcs$5 radius the current data also indicate that the central 
regions of this galaxy have a color that is superficially consistent with an 
old population. However, the color profile clearly indicates that there is a 
distinct nucleus that is bluer than the surroundings; hence, the 
central regions of NGC 2403 are not uniformly old.

	A color gradient is also seen in the central regions of NGC 247, 
although it is in the opposite sense to that in NGC 2403 
and M33. The $J-K$ color within the central 0$\farcs$5 
of NGC 247 is similar to that in M33, suggesting that the nuclear stellar 
contents of NGC 247 and M33 are similar, although this should be 
confirmed with observations of NGC 247 at higher angular resolutions. While the 
central light concentration of NGC 247 is bluer than the nucleus, 
it still has an SED that is consistent with an old simple stellar population.

	It is remarkable that the central regions of four of the nearest 
galaxies of morphological type Sbc (the Milky-Way) and Sc (M33, NGC 247, and 
NGC 2403) contain relatively young nuclear star clusters. It is also 
worth noting that the two nearest Sb systems, M31 and M81, show evidence 
for a young or intermediate-age nuclear population (Davidge 1997, Sil'Chenko et 
al. 1998, Davidge \& Courteau 1999) and/or nuclear activity (Filippenko \& 
Sargent 1988; Davidge \& Courteau 1999), although the nuclear population in 
M31 is older than what is seen near the centers of 
the Milky-Way and nearby Sc galaxies (Davidge 2001b). The high frequency of 
young nuclear populations among the closest galaxies is consistent with 
surveys of more distant systems. Of the 45 unbarred spirals with 
morphological types Sb and later observed by B\"{o}ker et al. (1999), 17 
show nuclear Pa$\alpha$ emission. Ho, Filippenko, \& Sargent find an even 
higher incidence of line emission from star formation in late-type spirals, 
although their data have a typical spatial resolution of $200 \times 400$ 
parsecs, and so some of this emission undoubtedly originates from 
the inner disk rather than the nucleus. 

	Unless one is willing to accept that the 
central regions of nearby late-type spiral galaxies are being
observed at fortuitously similar stages of a once-in-a-galaxy-lifetime 
occurence, then it appears that nuclear star formation is an episodic or 
continuous phenomenon in `normal' spiral galaxies, with a maximum time between 
star-forming events such that the nucleus can be easily detected with respect 
to the surroundings. An estimate for the time scale between significant 
nuclear star forming events can be obtained from the B\"{o}ker et al. 
(1999) survey if it is assumed that the Pa$\alpha$ emission is due to 
star formation, which is reasonable given that AGN and LINER emission 
tends to occur in galaxies earlier than type Sbc (Ho et al. 1997).
If line emission can be detected for $\sim 0.1$ Gyr after a burst of star 
formation, then the time between star-forming episodes is 0.1 Gyr $\times 
45/17 \sim 0.3$ Gyr.

	Serabyn \& Morris (1996) discuss mechanisms by which gas from the 
Galactic disk can be funneled into the central molecular zone (CMZ), which in 
turn fuels star formation in the central regions of the Galaxy. Observational 
support of a disk origin for the star-forming material near the GC comes from 
the flattened nature of the CMZ, which also lies along the Galactic Plane 
(Bally et al. 1988), and abundance studies of young stars near the GC (Ramirez 
et al. 2000), although these measurements are restricted at present to Fe 
lines, and the chemical mixtures in these stars, which will provide a signature 
of the origin of the material from which they formed, remain to be determined. 
Serabyn \& Morris (1996) argue that the central cusp in the Galactic bulge is 
sustained by either continuous or frequent bursts of star formation, 
and they note that a key test of this hypothesis is to determine whether 
young or intermediate-age nuclear populations are present in other galaxies. 
{\it The current data indicate that recent nuclear star formation in nearby 
galaxies is a common phenomenon, thus supporting the Serabyn \& 
Morris (1996) model of frequent or continuous central star formation.} 
We note further that the $J-K$ color of the NGC 2403 nucleus suggests that 
it is younger than the nuclei in the other Sc galaxies if the central 
extinctions in these systems are similar, and this is consistent 
with the relative ages of the inner disks, as infered from the resolved bright 
stellar content. While a larger sample of systems must be observed to 
establish a rigorous correlation, this is a tantalizing hint
that star formation in the nucleus and inner disk might have common triggers.

	Is nuclear star formation affected by, or related to, the presence of a 
super-massive central object? A connection might be expected 
since super-massive objects affect the central structural characteristics of 
galaxies (e.g. Faber et al. 1997), and impose demanding constraints on the 
conditions required to form stars near galaxy centers (e.g. Figer et al. 2000). 
Indeed, the correlation between the masses of super-massive objects and the 
velocity dispersion of their host bulges (Ferrarese \& Merritt 2000; Gebhardt 
et al. 2000) suggests that the formation and evolution of the central black 
hole and bulge are coupled (e.g. Haehnelt \& Kauffmann 2000). Nevertheless, 
Salucci et al. (2000) find that late-type spiral galaxies may not fall along 
the relation between black hole mass and bulge mass defined by early-type 
spirals and ellipticals, in the sense that the central objects in late-type 
galaxies are less massive than those in early-type galaxies having the same 
bulge mass. This could be a signature of morphology-related systematic 
differences in the evolutionary histories of bulges, and may explain the 
tendency for AGN to be found in early-type spirals (e.g. Ho et al. 1987). In 
any event, while the nuclei of M33 and the Galaxy have similar 
stellar contents (Paper 1), the GC contains a central compact object (Ghez 
et al. 1998), which is much more massive than that in M33 (e.g. 
Kormendy \& McClure 1993; Lauer et al. 1998). Thus, it appears that an 
extremely massive central object is not essential to trigger nuclear star 
formation, at least among intermediate and late-type spiral galaxies; rather, 
the results in this paper suggest that nuclear star formation in these systems 
is driven by non-local factors that also encompass the inner disk.

\vspace{0.3cm}
	It is a pleasure to thank an anonymous referee for comments that 
helped improve the paper.

\clearpage

\begin{table*}
\begin{center}
\begin{tabular}{lccccc}
\tableline\tableline
Galaxy & Type\tablenotemark{a} & B$_T$\tablenotemark{b} & A$_B$\tablenotemark{c} & $\mu_0$\tablenotemark{d} & M$_B$ \\
\tableline
NGC 247 & Sc(s)III-IV & 9.74 & 0.46 & 27.3 & --18.0 \\
NGC 2403 & Sc(s)III & 8.74 & 0.34 & 27.5 & --19.1 \\
M33 & Sc(s)II-III & 6.31 & 0.32 & 24.5 & --18.5 \\
\tableline
\end{tabular}
\end{center}
\tablenotetext{a}{From Sandage \& Tammann (1987)}
\tablenotetext{b}{From Table 1 of Pierce \& Tully (1992)}
\tablenotetext{c}{Sum of Internal and Galactic extinction from Pierce \& Tully (1992)}
\tablenotetext{d}{The distance moduli for NGC 2403 and M33 are based on 
Cepheids, using results from Freedman \& Madore (1988) and van den Bergh 
(1992). Lacking published observations of Cepheids in NGC 247, 
the distance modulus for this galaxy is based on the 
Tully-Fisher relation using integrated brightnesses and HI line 
widths from Pierce \& Tully (1992), and the calibration of Sakai et al. (2000).}
\caption{Galaxy Properties}
\end{table*}

\clearpage

\begin{center}
FIGURE CAPTIONS
\end{center}

\figcaption
[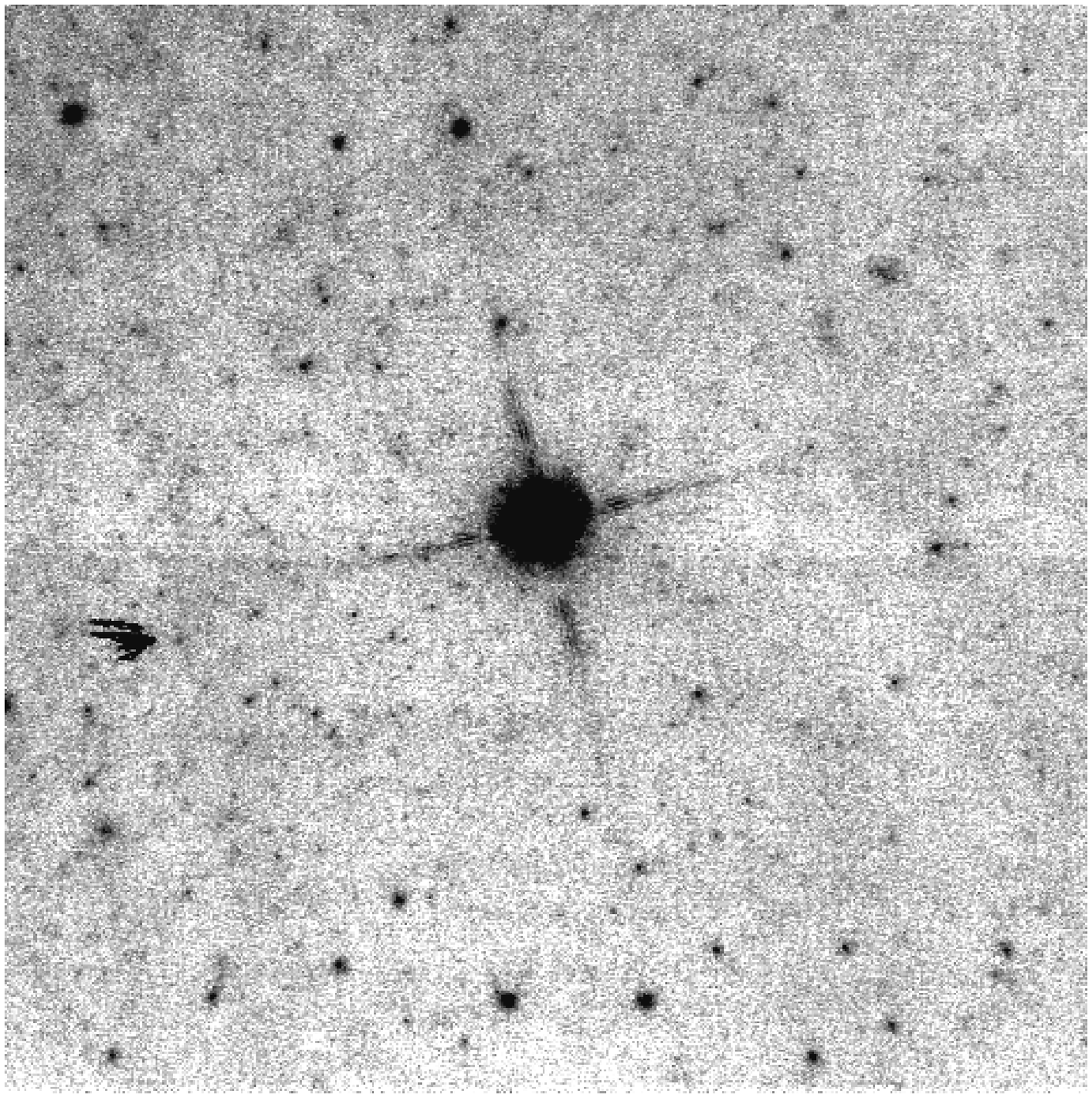]
{The final $Ks$ image of the NGC 2403 deep field. The image covers $34 \times 
34$ arcsec with North at the top and East to the left. The bright central 
source is the uncatalogued $K = 10$ star with $\alpha$ = 07:36:48 and $\delta$ 
= $+$65:35:41 (E2000) that was used as the AO guide star.}

\figcaption
[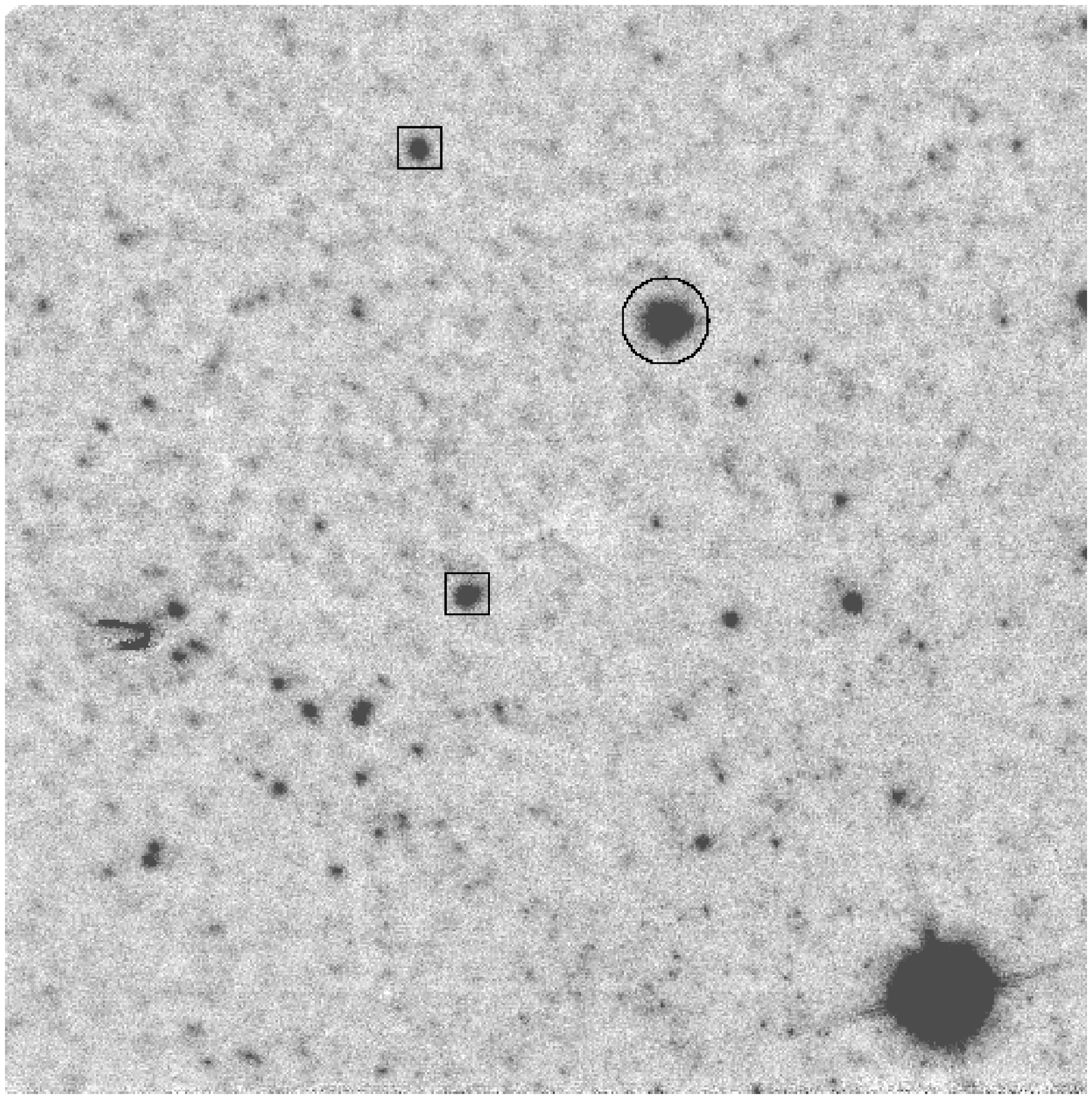]
{The final $Ks$ image of the NGC 2403 central field. 
The image covers $34 \times 34$ arcsec with North at the top and East to the 
left. The bright source in the lower right hand corner is the 
uncatalogued $K = 10$ star that was used as the AO reference beacon, 
and defines the center of the NGC 2403 deep field in Figure 1. The central 
light concentration of NGC 2403 is circled, while squares mark the two 
extended objects discussed in \S 3.2.}

\figcaption
[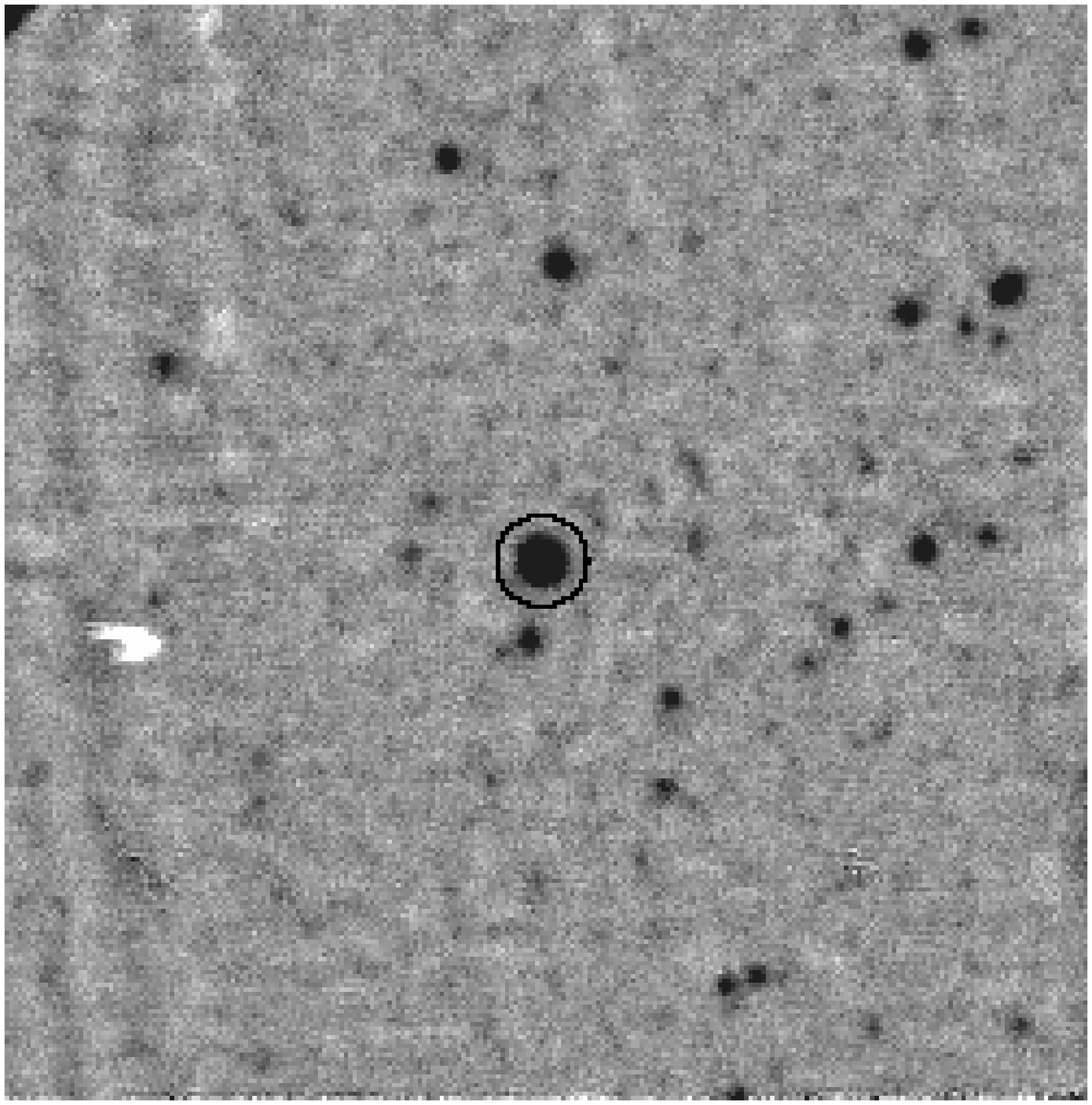]
{The final $Ks$ image of the NGC 247 central field. 
The image covers $34 \times 34$ arcsec with North at the top and East to the 
left. The central light concentration of NGC 247, which served as the reference 
source for AO guiding, is circled.}

\figcaption
[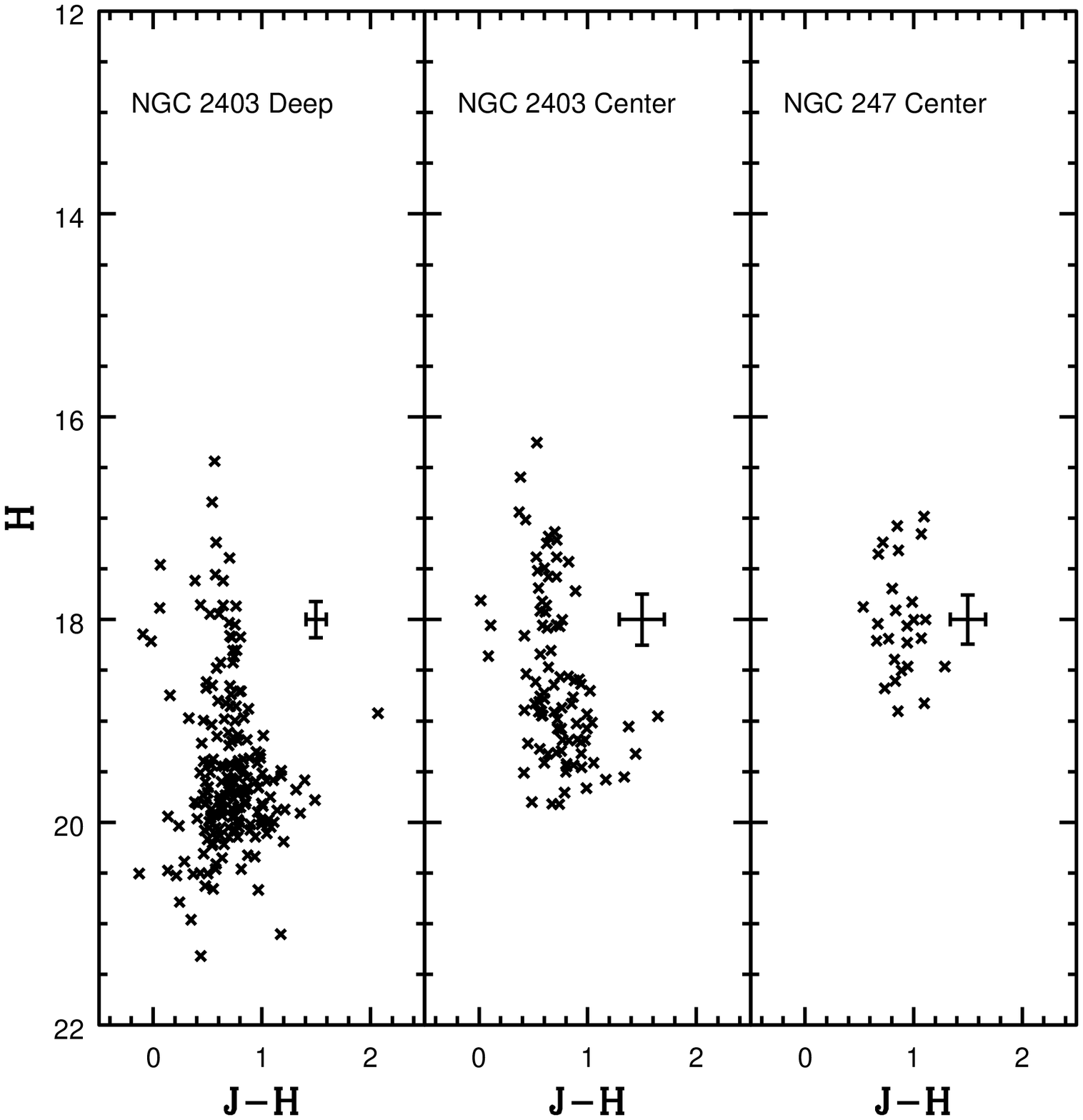]
{The $(H, J-H)$ CMDs of stars in the NGC 2403 and NGC 247 fields that 
were detected in all 3 filters. The error bars show the $1\sigma$ 
uncertainties predicted from artificial star experiments. The predicted 
uncertainties match the observed width of the red plume in each galaxy, 
indicating that the scatter in the CMDs is dominated by observational errors.
The stars with $J-H = 0$ in both NGC 2403 fields are likely early-type 
supergiants.}

\figcaption
[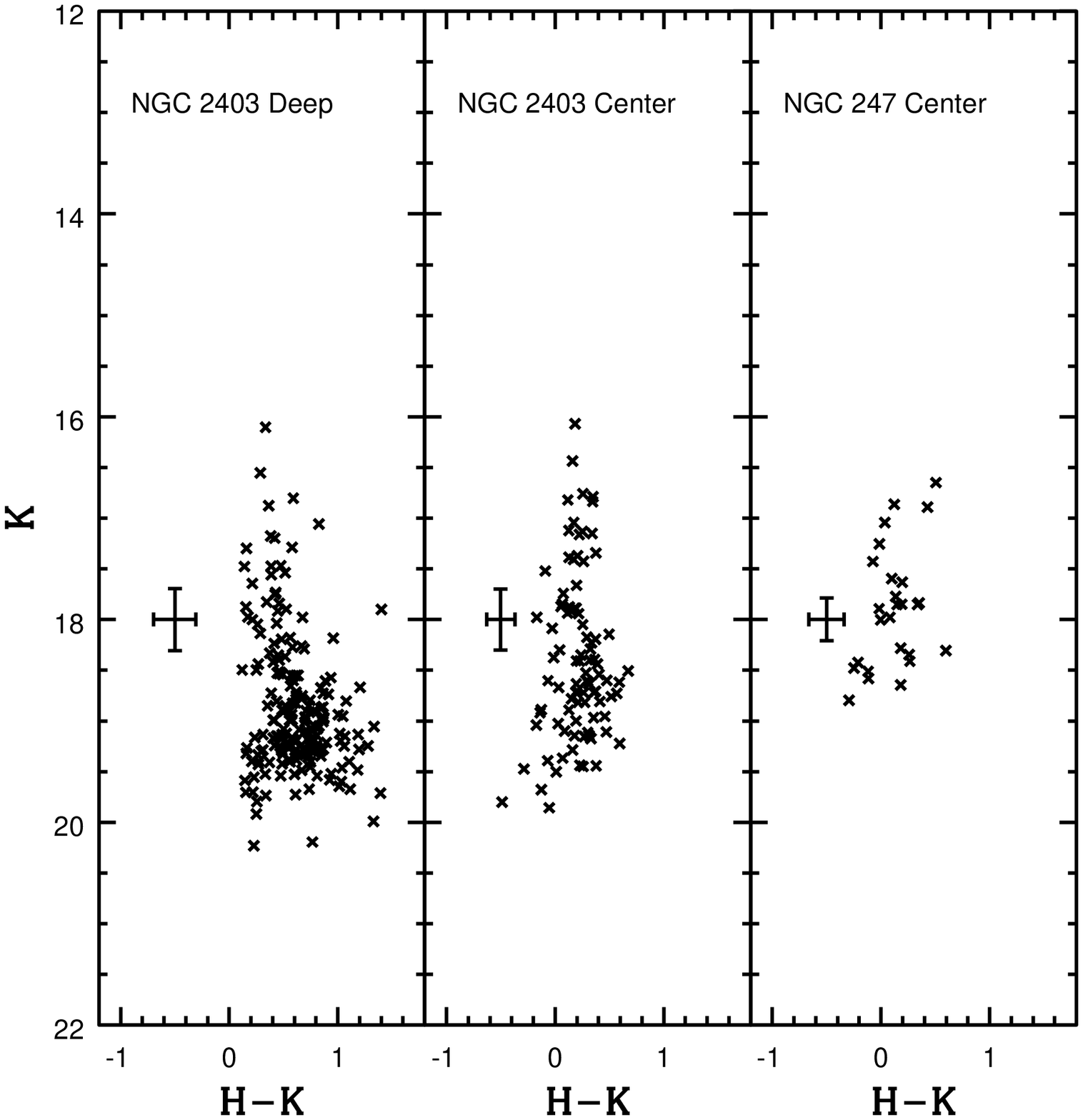]
{The $(K, H-K)$ CMDs of stars in the NGC 2403 and NGC 247 fields that 
were detected in all 3 filters. The error bars show the $1\sigma$ 
uncertainties predicted from artificial star experiments. 
The predicted uncertainties generally match the width of 
the red plume in each galaxy, indicating that the scatter in the CMDs is 
dominated by observational errors.}

\figcaption
[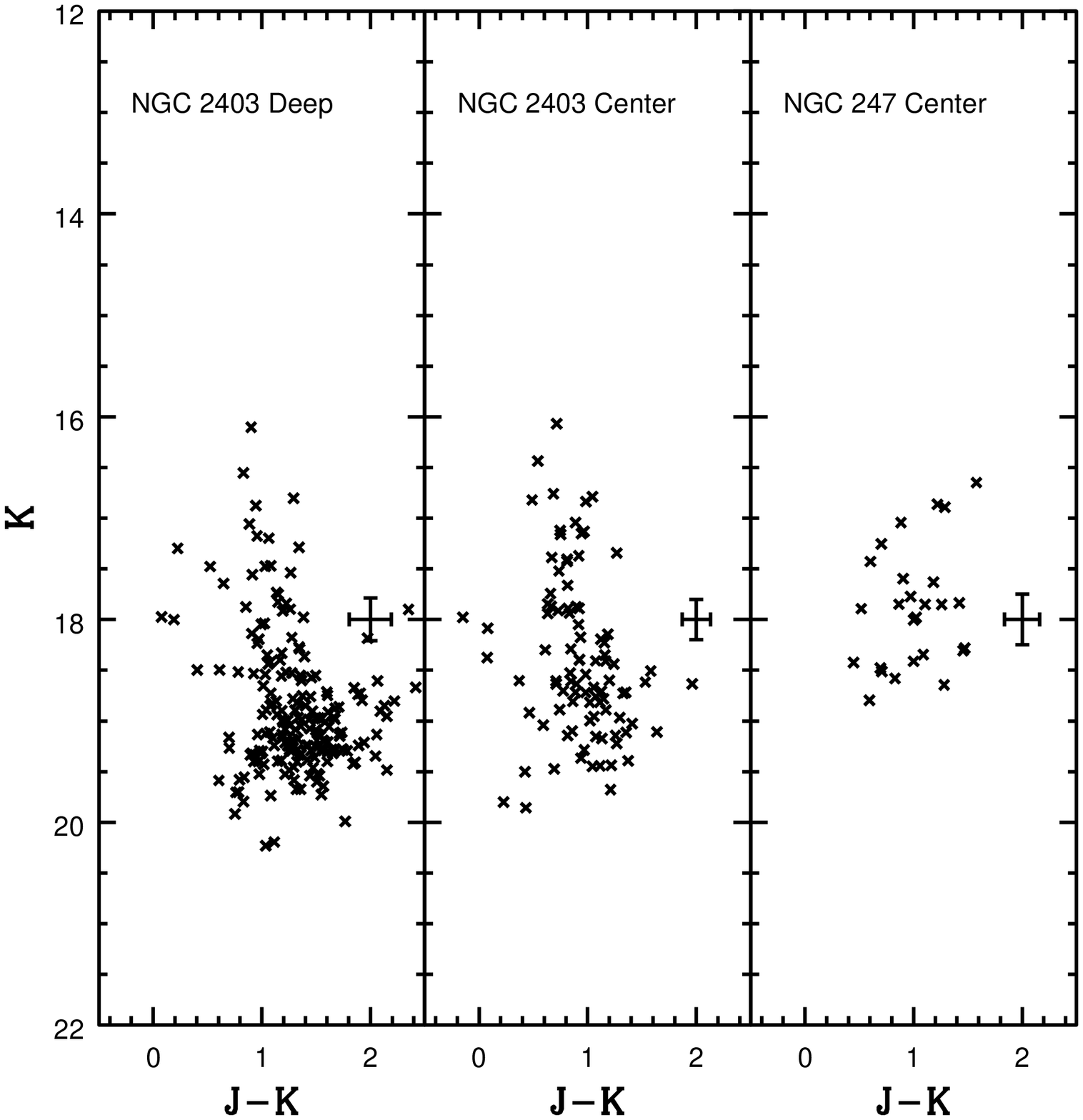]
{The $(K, J-K)$ CMDs of stars in the NGC 2403 and NGC 247 fields that 
were detected in all 3 filters. The error bars show the $1\sigma$ 
uncertainties predicted from artificial star experiments. The predicted 
uncertainties match the width of the red plume in each galaxy, 
indicating that the scatter in the CMDs is dominated by 
observational errors. The stars with $J-K = 0$ in both 
NGC 2403 fields are early-type supergiants.}

\figcaption
[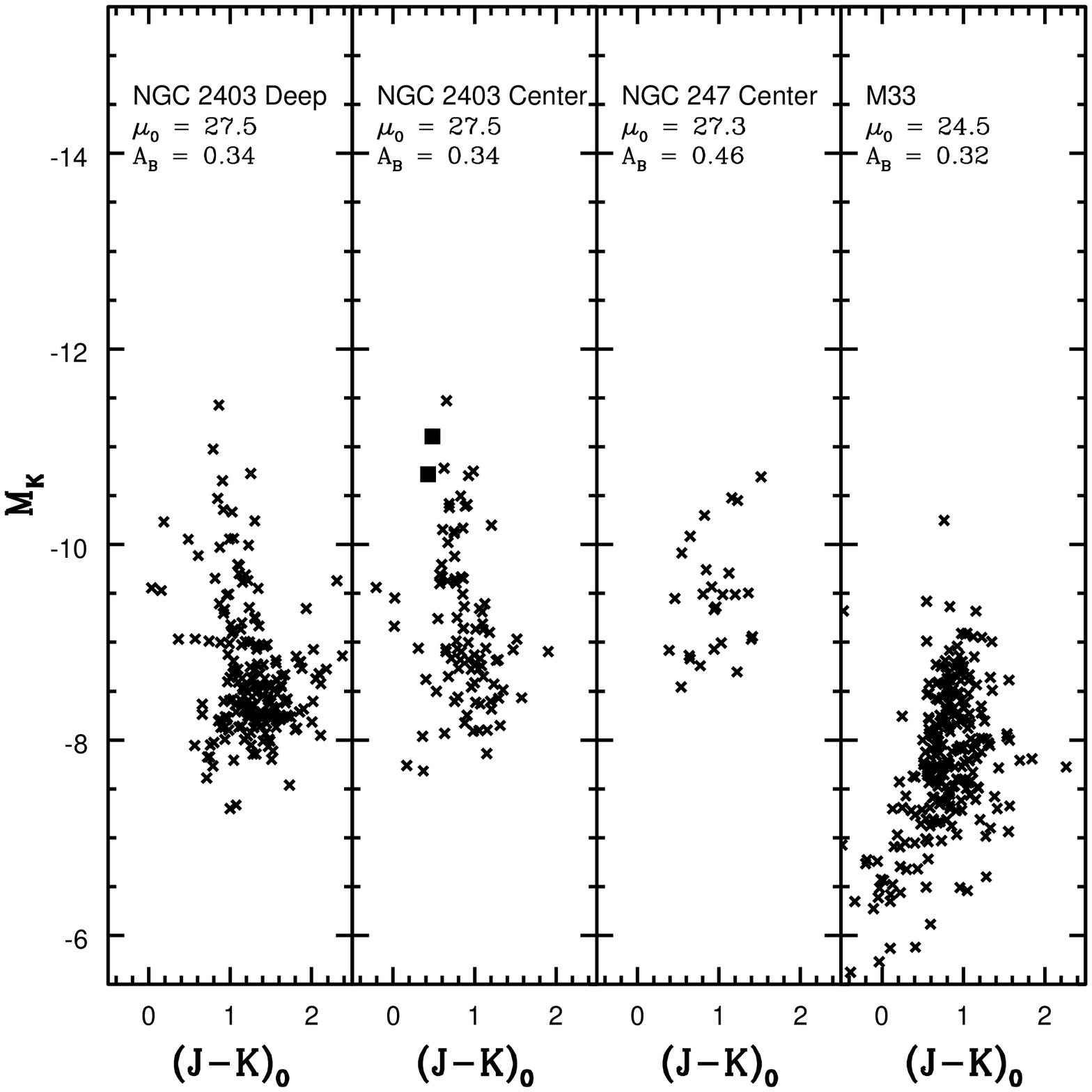]
{The $(M_K, J-K)$ CMDs of the NGC 2403 deep field, the central fields 
of NGC 2403 and NGC 247, and the central regions of M33. The M33 CMD was 
obtained from the data discussed in Paper 1, after these were processed 
to simulate moving this galaxy to the distance of NGC 2403 and then 
observing with the same angular resolution as the NGC 2403 central field. 
The filled squares in the NGC 2403 central field CMD mark sources 
that are non-stellar in appearance. The adopted distance moduli and 
extinctions for each field are shown. Note the conspicuous differences in the 
bright stellar contents of the three galaxies.}

\figcaption
[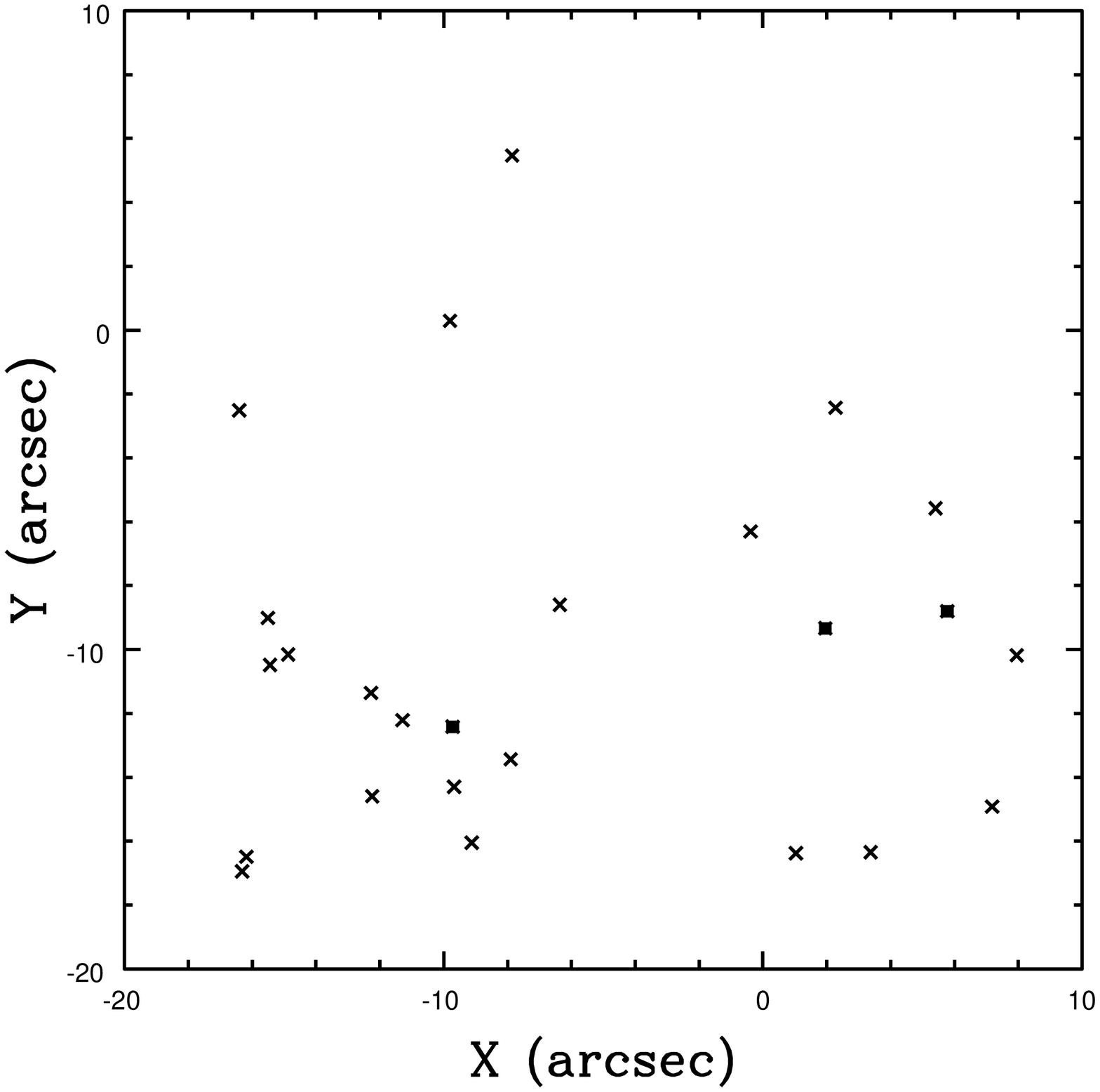]
{The spatial distribution of RSGs within two magnitudes of the peak $K$ 
brightness in the NGC 2403 central field. Stars with M$_K$ between $-9.5$ and 
$-10.5$ are plotted as crosses, while stars with M$_K \leq -10.5$ are shown as 
filled squares. $X$ and $Y$ are offsets, in arcsec, from the center of 
NGC 2403, with North at the top, and East to the left. Note that the vast 
majority of these bright stars are located to the South of the nucleus.}

\figcaption
[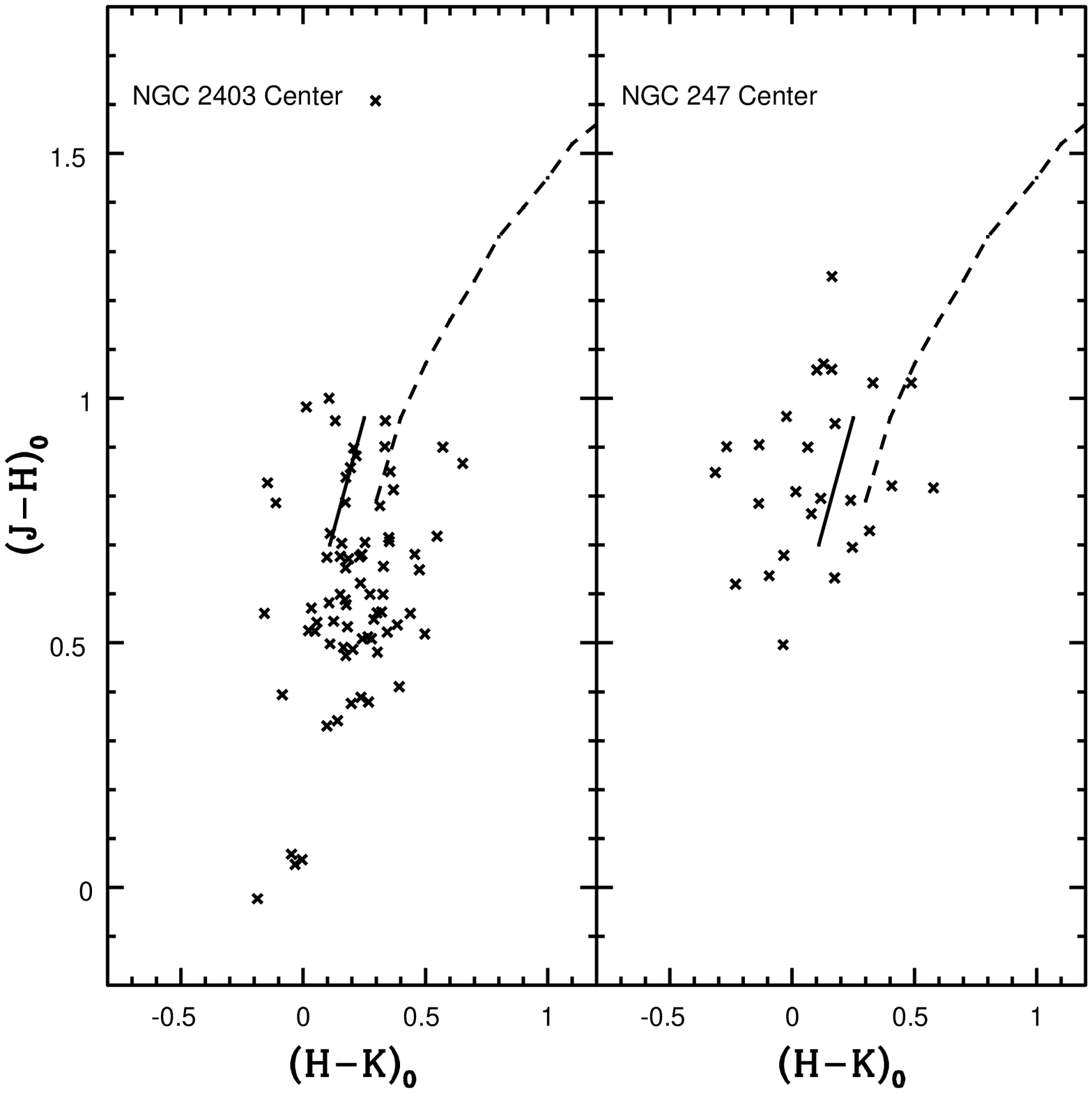]
{The $(J-H, H-K)$ two color diagrams of stars in the NGC 2403 
and NGC 247 central fields. The dashed line is the locus of LPVs in the 
LMC and SMC, defined from observations published by Wood et al. (1983, 
1985), while the solid line is the locus of Milky-Way RSGs, established from 
data published by Elias et al. (1985).}

\figcaption
[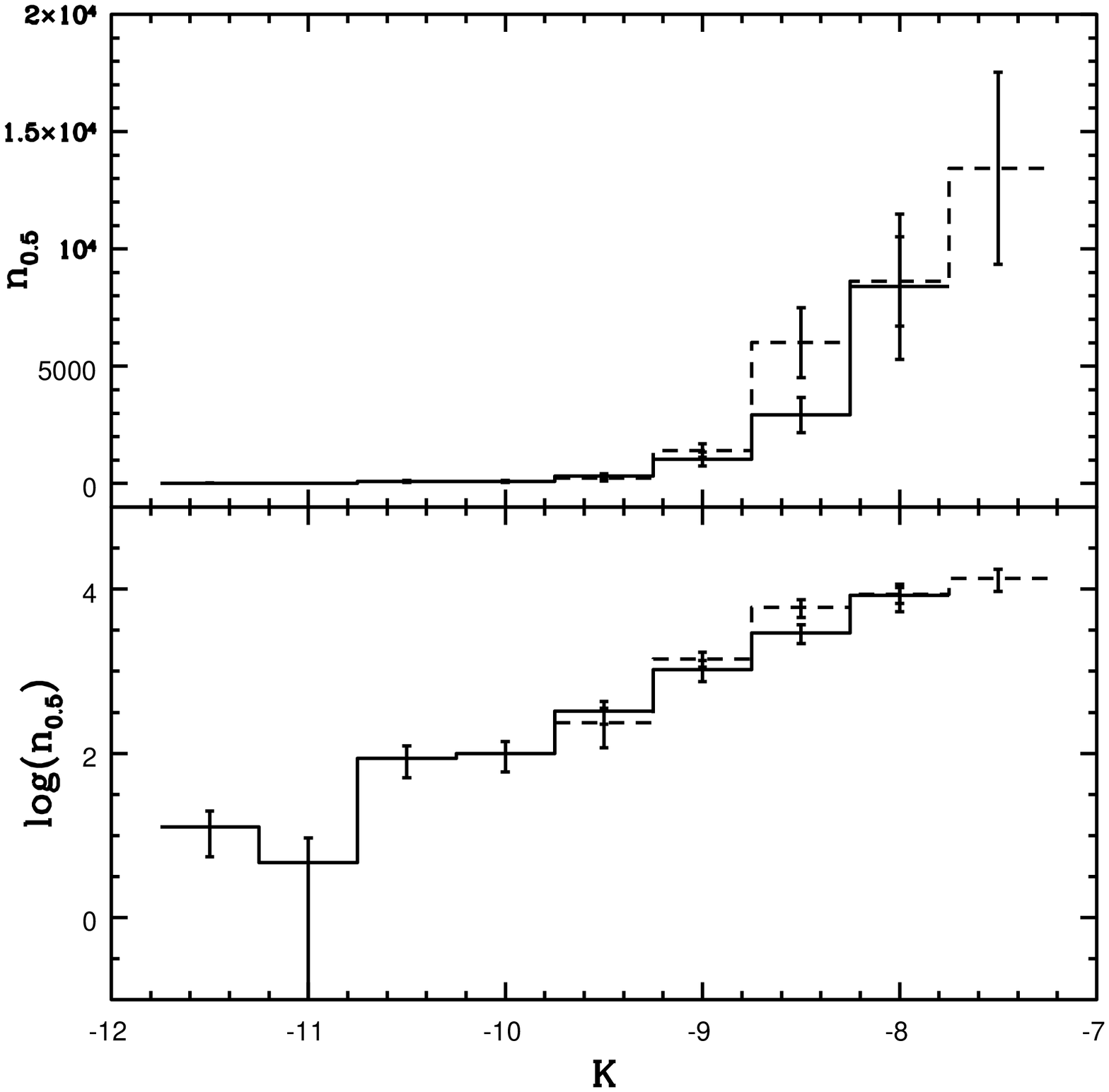]
{The completeness-corrected M$_K$ LFs of stars near the centers of NGC 2403 
(solid line) and M33 (dashed line), where n$_{0.5}$ is the number of stars per 
0.5 mag interval per kpc$^2$. The error bars show uncertainties due to 
counting statistics and completeness corrections. The NGC 2403 LF, which 
was constructed by combining data in the deep and central fields using the 
procedure described in the text, has been scaled to match the $r-$band surface 
brightness in the M33 field using the light profiles measured 
by Kent (1987a, b). The LFs of NGC 2403 and M33 are in excellent 
agreement between M$_K = -9.5$ and $-8.0$, indicating that the star-forming 
histories of the inner disks of these galaxies have been similar over time 
scales probed by bright AGB stars ($\sim 1$ Gyr).} 

\figcaption
[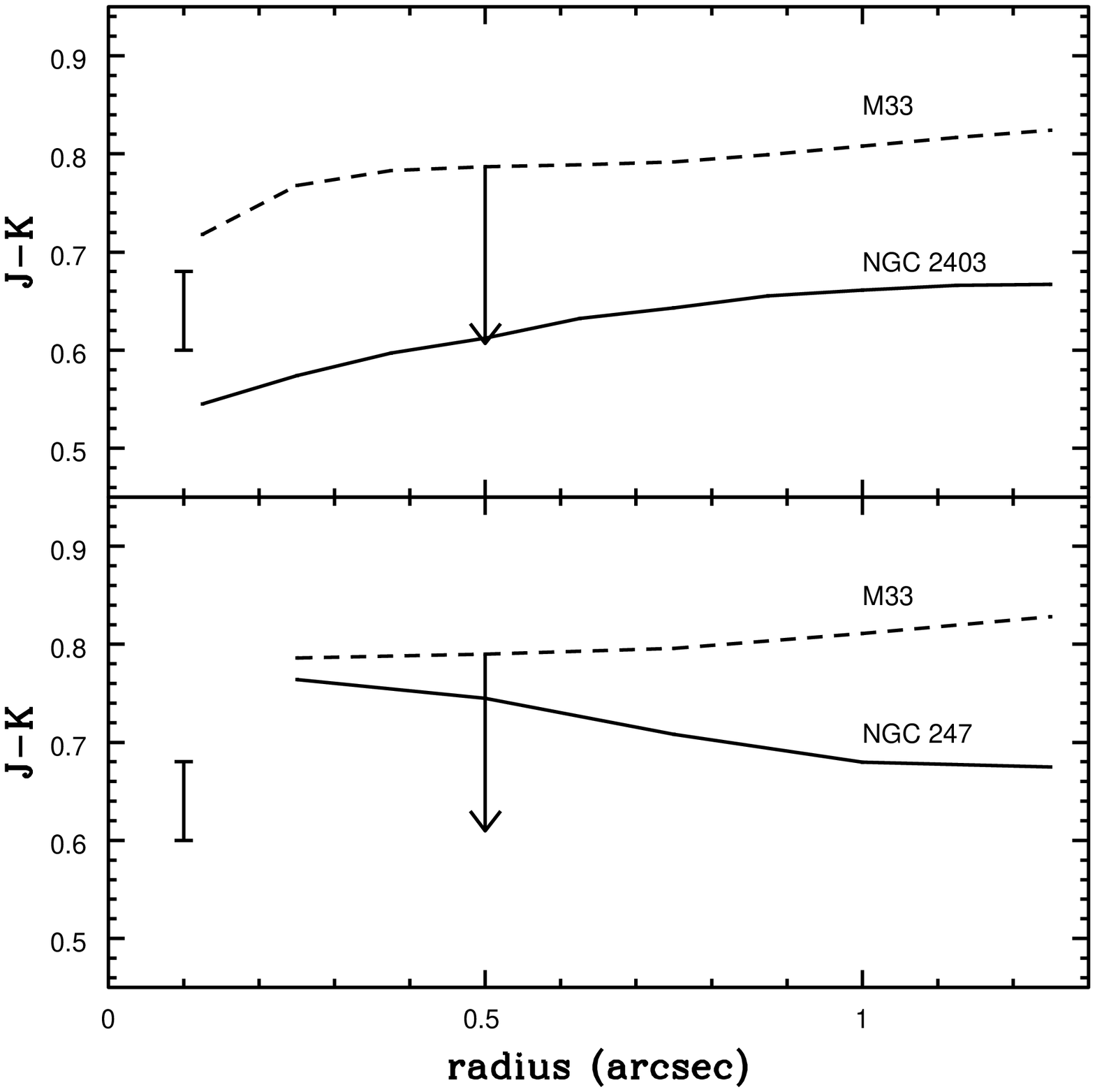]
{The integrated $J-K$ color profiles of 
NGC 2403 and NGC 247. The dashed lines show the corresponding color curves 
for M33 measured from data processed to simulate the appearance 
of the central field of this galaxy if viewed at the same distance and with 
the same angular resolution as NGC 2403 and NGC 247. The bluest nuclear $J-K$ 
color occurs in NGC 2403, and the reddest is in M33; this is broadly consistent 
with the age sequence predicted from the brightest red stars in the inner 
disks of these galaxies. The $1\sigma$ error bar showing the uncertainty in the 
photometric calibration, which is the dominant source of random error in the 
integrated color measurements, is also shown. The arrow shows the result 
of adopting A$_V = 1.4$ mag for the nucleus of M33, as measured by 
Gordon et al. (1999). Note that the M33 and NGC 2403 curves in the upper panel 
differ at roughly the $5\sigma$ level if these systems have similar amounts 
of extinction in their central regions. However, if the extinction in the 
nucleus of NGC 2403 is significantly less than that in M33 then the nuclei 
of these galaxies may have similar intrinsic colors.}

\figcaption
[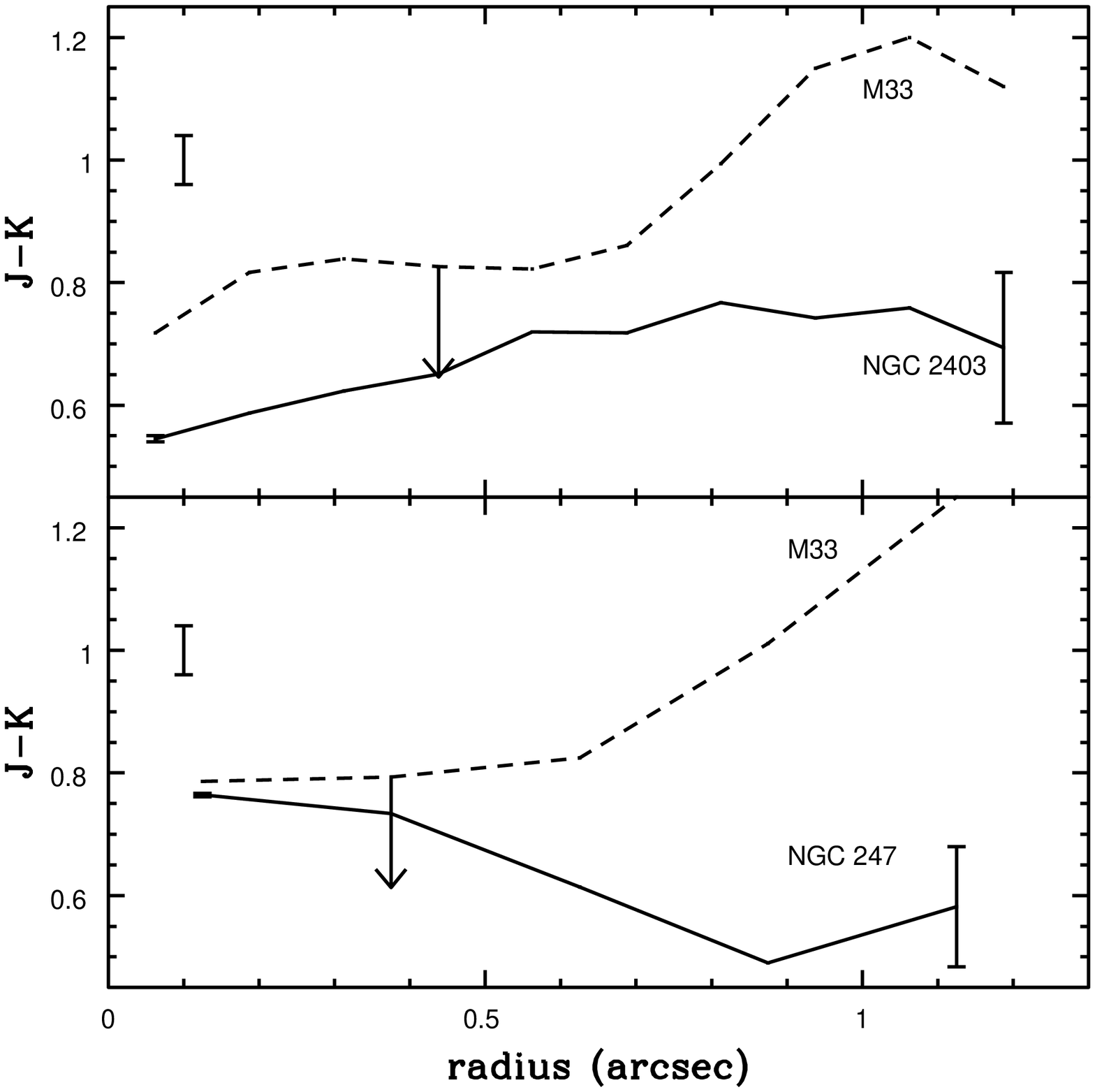]
{The $J-K$ colors measured in concentric circular annuli for 
NGC 2403 and NGC 247. The dashed lines show the corresponding color curves 
for M33 measured from data processed to simulate the appearance 
of the central field of this galaxy if viewed at the same distance and with the 
same angular resolution as NGC 2403 and NGC 247. The bluest nuclear $J-K$ 
color occurs in NGC 2403, while the reddest is in M33; this is broadly 
consistent with the age sequence infered from bright red stars in 
the inner disks of these galaxies. The error bars on the end points of the NGC 
2403 and NGC 247 curves show the $1\sigma$ uncertainties 
due to photon statistics and sky subtraction. The $1\sigma$ 
error bar in the photometric calibration, which is the dominant source of 
random error in the M33 data, is shown near the upper left hand corner of each 
panel. The arrow shows the result of adopting A$_V = 1.4$ mag for the nucleus 
of M33, as measured by Gordon et al. (1999).}

\figcaption
[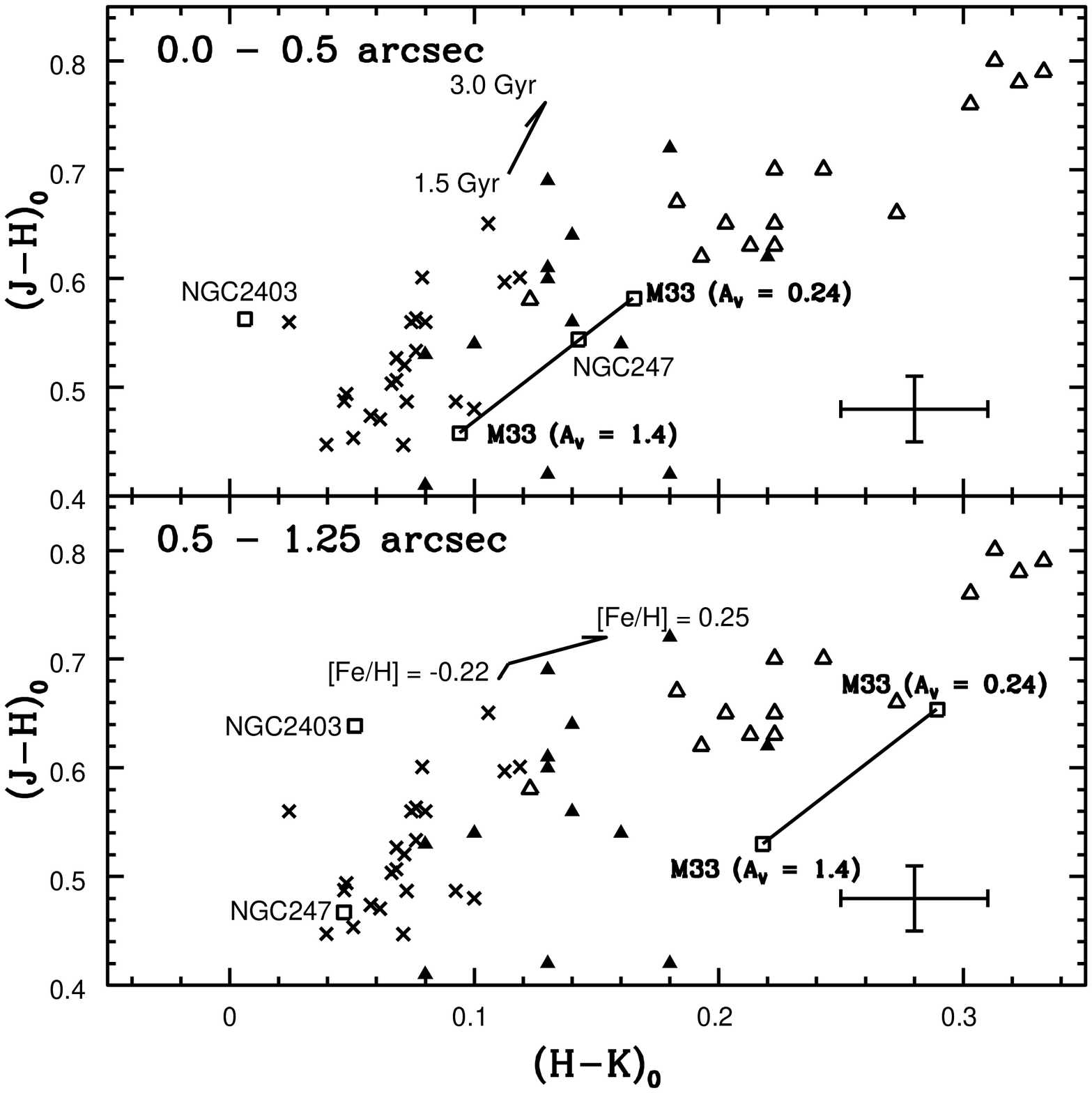]
{The $(J-H, H-K)$ TCD for the central regions of NGC 247, NGC 2403, and M33. 
The M33 measurements were made from images that were processed to match the 
distance and angular resolution of the NGC 2403 observations; these do not 
change greatly if measured from images processed to match the image 
quality of the NGC 247 data. Plotted are the luminosity-weighted colors 
within $0\farcs$5 (top panel) and between $0\farcs$5 and 1$\farcs$25 
(lower panel) of the center of each galaxy. Two points are plotted 
for M33 to show the sensitivity to extinction: one point assumes 
A$_V = 0.24$, as estimated by Pierce \& Tully 
(1992), while the other assumes A$_V = 1.4$, based on the fit to the 
nuclear SED made by Gordon et al. (1999). The error bars show the uncertainty 
in the photometric calibration. Also plotted are data for (1) Galactic globular 
clusters, as listed in Table 5A of Brodie \& Huchra (1990) and corrected for 
reddening using the $E(B-V)$ values listed in that table (crosses), (2) 
Magellanic Cloud SWB Type 1 and 2 clusters from Tables 
4 and 5 of Persson et al. (1983) (filled triangles), and (3) 
the central regions of Sc galaxies, as listed in Table 4 of Frogel (1985). 
Frogel (1985) only corrected for foreground 
extinction, and the points plotted in Figure 13 have been 
corrected for internal disk extinction using the relation derived by Tully 
\& Fouqu\'{e} (1985) assuming $i = 45^{o}$. The arrow in the top panel 
connects the 1.5 Gyr and 3.0 Gyr [Fe/H] = 0 models from Worthey (1994), 
while the arrow in the lower panel connects the [Fe/H] = -0.22, [Fe/H] = 0, 
and [Fe/H] = 0.25 Worthey (1994) models with an age of 1.5 Gyr.}

\figcaption
[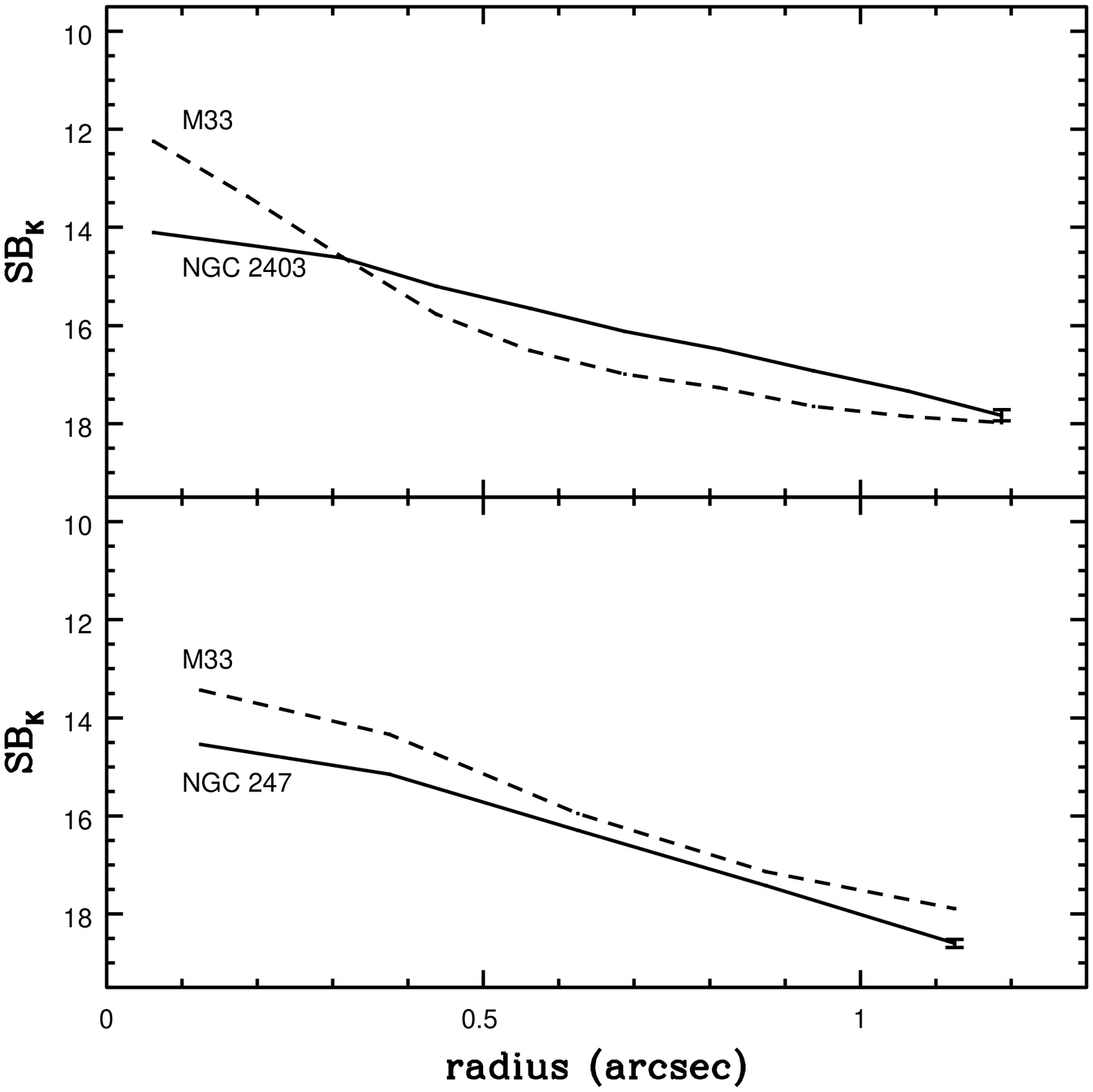]
{The $K-$band surface brightnesses of NGC 2403, NGC 247, and M33 as 
measured in concentric circular apertures. The 
dashed lines show the surface brightness profiles of M33 measured from data 
processed to simulate the appearance of the central field of this galaxy if 
viewed at the same distance and with the same angular resolution as NGC 247 and 
NGC 2403. Note that the mean central surface brightness of M33 
is significantly higher than in the other galaxies. The error 
bars at the end points of the NGC 2403 and NGC 247 curves show the $1\sigma$ 
uncertainties due to photon statistics and sky subtraction.}
\end{document}